\documentclass{amsart}

\usepackage{amssymb}
\usepackage{graphicx}
\usepackage{epstopdf}
\usepackage{amscd}
\usepackage{graphics}
\newtheorem{Theorem}{Theorem}[section]
\newtheorem{Proposition}{Proposition}[section]

\newtheorem{Lemma}{Lemma}[section]

\def\proof{\par{\it Proof}. \ignorespaces}
\def\endproof{{\ \vbox{\hrule\hbox{%
     \vrule height1.3ex\hskip0.8ex\vrule}\hrule }}\par}
\newenvironment{Proof}{\proof}{\endproof}

\theoremstyle{definition}
\newtheorem{Definition}[Theorem]{Definition}
\newtheorem{Example}[Theorem]{Example}

\theoremstyle{remark}
\newtheorem{Remark}[Theorem]{Remark}

\numberwithin{equation}{section}

\begin{document}


\title{$N$-soliton solutions to the DKP equation and
 Weyl group actions}

\author{Yuji Kodama$^*$}
\address{Department of Mathematics, Ohio State University, Columbus,
OH 43210}
\email{kodama@math.ohio-state.edu}
\author{Ken-ichi Maruno$^{\dagger}$}
\address{Faculty of Mathematics, Kyushu University, Fukuoka, 812-8581, Japan}
\thanks{$^*$Partially
supported by NSF grant DMS0404931.
$^{\dagger}$ Supported by COE program at Faculty of Mathematics, Kyushu University.}

\email{maruno@math.kyushu-u.ac.jp}


\thispagestyle{empty}

\pagenumbering{arabic}
\setcounter{page}{1}

\begin{abstract}
We study soliton solutions to the DKP equation
which is defined by the Hirota bilinear form,
\[
\left\{
\begin{array}{llll}
(-4D_xD_t+D_x^4+3D_y^2)\,\tau_n\cdot\tau_n=24\tau_{n-1}\tau_{n+1}\,,\\
(2D_t+D_x^3\mp 3D_xD_y)\,\tau_{n\pm 1}\cdot\tau_n=0\\
\end{array}\right. \quad n=1,2,\ldots.
\]
where $\tau_0=1$.  The $\tau$-functions $\tau_n$ are given by
the pfaffians of certain skew-symmetric matrix.
We identify one-soliton solution as an element of the Weyl group of D-type, and
discuss a general structure of the interaction patterns among the solitons.
Soliton solutions are characterized by $4N\times 4N$ skew-symmetric constant
matrix which we call the $B$-matrices. We then find that one can have
$M$-soliton solutions with $M$ being any number from $N$ to $2N-1$ for
some of the $4N\times 4N$ $B$-matrices having only $2N$ nonzero
entries in the upper triangular part (the number of solitons obtained from those $B$-matrices
was previously expected to be just $N$).
\end{abstract}

\maketitle

\section{Introduction}
We study soliton solutions of the DKP equation \cite{adler:99,hirota:91,kac:98},
\begin{equation}\label{dkptau}
\left\{
\begin{array}{llll}
(-4D_xD_t+D_x^4+3D_y^2)\,\tau_n\cdot\tau_n=24\tau_{n-1}\tau_{n+1}\,,\\
{}\\
(2D_t+D_x^3\mp 3D_xD_y)\,\tau_{n\pm 1}\cdot\tau_n=0\\
\end{array}\right. \quad n=1,2\ldots,
\end{equation}
where $\tau_0=1$, and $D_x, D_y$ and $D_t$ are the usual Hirota derivatives.
For each $n$,
the variables $u=2( \ln\tau_n)_{xx}$ and $v^{\pm}=\tau_{n\pm 1}/\tau_n$
define the coupled KP equation (see for example  \cite{hirota:91}),
\[\left\{\begin{array}{llll}
(-4u_t+u_{xxx}+6uu_x)_x+3u_{yy}=24(v^+v^-)_{xx}\,,\\
{}\\
2v^{\pm}_t+v^{\pm}_{xxx}+3uv^{\pm}_x\mp 3(v^{\pm}_{xy}+v^{\pm}\int^xu_y\, dx)=0\,.
\end{array}\right.
\]
This set of equations admits a class of particular solutions, called soliton solutions
similar to those of the KP equation (see for example \cite{isojima:02}), and in this paper
we describe some properties of those solutions.
\begin{Remark}
The DKP equation was given as a first member of the DKP hierarchy in \cite{jimbo:83}.
In \cite{hirota:91}, the coupled KP equation was introduced as an extension of the KP equation
whose solutions are given by the pfaffian form.
The DKP equation was rediscovered as the Pfaff lattice describing the partition
function of a skew-symmetric matrix model in \cite{adler:99, kakei:99},
and was also found as an orbit of some infinite-dimensional Clifford group
action in \cite{kac:98}. 
\end{Remark}

The $\tau_n$ functions are given by the pfaffians of $2n\times 2n$ skew-symmetric matrices ${\mathcal Q}_n$
whose entries are denoted by $Q_{i,j}$ for $1\le i<j\le 2n$ with $Q_{j,i}=-Q_{i,j}$ ,
\begin{equation}\label{tau}
\tau_n={\rm Pf}({\mathcal Q_n})=\sum_{\smaller{\begin{matrix}1=i_1<\ldots<i_n< 2n\\ i_k<j_k,~k=1,\ldots,n
\end{matrix}}} \sigma(i_1,j_1,\ldots,i_{n},j_n)Q_{i_1,j_1}Q_{i_2,j_2}\cdots Q_{i_n,j_n}
\end{equation}
Each coefficient $\sigma(i_1,j_1,\ldots,i_n,j_n)$ gives a sign corresponding to the parity of the permutation,
\[
\sigma:={\rm sign}\begin{pmatrix}
1 & 2 & \cdots & 2n-1&2n\\
i_1&j_1&\cdots&i_n&j_n
\end{pmatrix}
\]
Here the elements $Q_{i,j}$ satisfy
\[
\frac{\partial}{\partial t_k}Q_{i,j}=Q_{i+k,j}+Q_{i,j+k}\,,
\quad k=1,2,\ldots
\]
where $t_1=x,t_2=y,t_3=t$ and others are the symmetry parameters.
A realization of $Q_{i,j}$ is given by \cite{hirota:91},
\[
Q_{i,j}=\left|\begin{matrix}\phi^{(i-1)} &\phi^{(j-1)} \\ \psi^{(i-1)} & \psi^{(j-1)}\end{matrix}\right|\,,
\quad {\rm for}\quad i<j
\]
where $\phi^{(k)}=\partial^k\phi/\partial x^k$ (the same for $\psi^{(k)}$) and the functions
$\phi$ and $\psi$ satisfy the same equations for $x$ and $t$ evolutions,
\[
\frac{\partial \phi}{\partial y}=\frac{\partial^2 \phi}{\partial x^2},\quad \quad
\frac{\partial\phi}{\partial t}=\frac{\partial^3 \phi}{\partial x^3}\,.
\]
For an example of finite dimensional solution, we consider
\begin{equation}\label{phi}
\phi(x,y,t)=\sum_{m=1}^M a_mE_m(x,y,t),\quad \psi(x,y,t)=\sum_{m=1}^Mb_mE_m(x,y,t)\,,
\end{equation}
with some constants $a_m,b_m$ for $m=1,\ldots,M$. The function $E_m(x,y,t)$ is the exponential
function,
\[
E_m(x,y,t):=e^{\theta_m}\,\quad {\rm with}\quad \theta_m(x,y,t)=p_mx+p_m^2y+p_m^3t+\theta_m^0\,,
\]
where $p_m$ and $\theta_m^0$ are arbitrary constants, and throughout this paper we assume
the parameters ${\bf p}:=(p_1,p_2,\ldots,p_M)$ to be ordered as,
\begin{equation}\label{orderp}
p_1<p_2<\cdots <p_M\,.
\end{equation}
With those $\phi$ and $\psi$, the elements $Q_{i,j},~1\le i<j\le 2n$ become
\begin{equation}\label{Qij}\begin{array}{llll}
Q_{i,j}&=&\displaystyle{\sum_{1\le k<l\le M} }b_{k,l}\left|\begin{matrix}
E_k^{(i-1)} & E_k^{(j-1)}\\
E_l^{(i-1)} &E_l^{(j-1)}
\end{matrix}\right| \\
{}\\
&=&\displaystyle{\sum_{1\le k<l\le M}}b_{k,l}(p_kp_l)^{i-1}(p_l^{j-i}-p_k^{j-i})E_{k,l}\,,
\end{array}
\end{equation}
where $b_{k,l}=a_kb_l-a_lb_k$, and $E_{k,l}:=E_kE_l=\exp(\theta_k+\theta_l)$.

As a generalization of this form of $Q_{i,j}$, we consider an arbitrary
 $M\times M$ skew-symmetric matrix $B=(b_{k,l})_{1\le k,l\le M}$.
We then note that
the $2n\times 2n$ matrix ${\mathcal Q}_n$ in the $\tau$-function $\tau_n$ can be
expressed as
\begin{equation}\label{Q}
{\mathcal Q}_n(x,y,t)={\mathcal E}_n(x,y,t){B}{\mathcal E}_n(x,y,t)^T\,
\end{equation}
where ${\mathcal E}_n$ is a $2n\times M$ matrix with its transpose ${\mathcal E}_n^T$, and it is given by
\[
{\mathcal E}_n=\begin{pmatrix}
E_1&E_2&\cdots & E_{M}\\
\vdots&\vdots&\ddots&\vdots\\
E_1^{(2n-1)}&E_2^{(2n-1)}&\cdots&E_{M}^{(2n-1)}
\end{pmatrix}\,.
\]
In this paper, we discuss a classification problem of several soliton solutions given by (\ref{Q})
in terms of the $B$-matrix. Note here that $M>2$ for $n\ge 1$, since the case with $M=2$
gives a trivial solution, i.e. $\tau_1=Q_{1,2}=b_{1,2}(p_2-p_1)E_1E_2$ and $u=2(\ln \tau_1)_{xx}=0$.
Also note that one needs $M> 2n$ for $u=2(\ln\tau_n)_{xx}\ne 0$.

One should also note that if $\tau_{n+1}=0$ (i.e. $v^+=0$), the corresponding solution 
satisfies the KP equation. This condition can be of course obtained from the
structure of the $B$-matrix. For example, if we take $M=3$, then $\tau_2$ vanishes
identically (the size of the pfaffian is $4\times 4$, but the independent exponentials
are three or less). This implies that $\tau_1$ with $M=3$ gives a solution of the KP
equation, and it gives either one KP soliton solution or a resonant Y-shape KP soliton: With
a $3\times 3$ $B$-matrix, we have
\[
\tau_1=Q_{1,2}=b_{1,2}(p_2-p_1)E_1E_2+b_{1,3}(p_3-p_1)E_1E_3+b_{2,3}(p_3-p_2)E_2E_3\,.
\]
 The function $u=2( \ln\tau_1)_{xx}$ gives one KP soliton solution if one of
 $b_{i,j}$ is zero (with others being positive), and Y-shape KP soliton if all $b_{i,j}$ are positive.
 For example, with $b_{1,2}=0$, we have $\tau_1=(b_{1,3}(p_3-p_1)E_1+b_{2,3}(p_3-p_2)E_2)E_3$
 which leads to
 \begin{equation}\label{Asoliton}
 w(x,y,t):=\frac{\partial }{\partial x}\ln \tau_1=p_3+\frac{1}{2}(p_1+p_2)+\frac{1}{2}(p_2-p_1){\rm tanh}\frac{1}{2}(\theta_2-\theta_1).
 \end{equation}
 The asymptotic values of $w$ then take
 \[
 w(x,y,t)\to\left\{\begin{array}{lll}
 p_1+p_3, \quad &{\rm for}~~& x\to -\infty\\
 p_2+p_3,\quad &{\rm for}~~ & x\to \infty
 \end{array}\right.
 \]
 Note here that the one soliton exchange the asymptotic values of $(1,3):=p_1+p_3$ and $(2,3):=p_2+p_3$,
 that is, this one soliton permutes the numbers, $(1)\leftrightarrow (2)$. We then label one soliton of the KP
 equation as an element of the permutation group $W$, in this case $W=S_3$, the symmetry
 group of order 3.
 We denote the one soliton solution (\ref{Asoliton}) by $[1:2]$, and call this type of soliton A-soliton
 (``A'' stands for type A Lie algebra which is the underlying symmetry algebra for KP equation).
 In general, we denote one A-soliton by $[i:j]$, if the function $w$ exchanges $p_i\leftrightarrow p_j$
 with $p_i<p_j$.
 We sometime identify $[i:j]$ as an element of the symmetry group.

 A generic solution of the DKP equation is then obtained for $M\ge 4$. In particular,
 we obtain one soliton solution in the case $M=4$ with the $B$-matrix having
 just two nonzero elements in the upper triangular part. For example, we consider the $B$-matrix having $b_{1,2}=1, b_{3,4}=1$ and all others $b_{i,j}=0$ for $i<j$. Then
\[
\tau_1=Q_{1,2}=(p_{2}-p_{1})E_{1}E_{2}+(p_{4}-p_{3})E_{3}E_{4}\,.
\]
This gives
\begin{equation}\label{oneDsol}
w(x,y,t):=\frac{\partial}{\partial x}\ln\tau_1=\frac{1}{2}\sum_{k=1}^4p_{k}+\frac{1}{2}(p_{3}+p_4-p_1-p_2)
{\rm tanh}\frac{1}{2}(\theta_{34}-\theta_{12})\,
\end{equation}
with $\theta_{ij}=\theta_i+\theta_j+\ln|p_i-p_j|$. 
For each asymptotic of $x\to\pm\infty$, one of the exponential terms in the $\tau$-function
becomes dominant. In the case of (\ref{oneDsol}), we have
\[
w(x,y,t)\to \left\{\begin{array}{llll}
p_{1}+p_{2},\quad {\rm as}~ x\to-\infty\,,\\
p_{3}+p_{4},\quad {\rm as}~ x\to\infty\,.
\end{array}
\right.
\]
Thus, this one D-soliton exchanges $(1,2) \leftrightarrow (3,4)$ with the values of $w$ where
$(i,j):=p_i+p_j$.
This indicates a Weyl-action of $D$-type: For example, $\pi_{1,2}\in W^D$
 may be expressed as
\[
\pi_{1,2}\cdot (i_1,i_2:i_{\bar 1},i_{\bar 2})=(i_2,i_1:i_{\bar 2},i_{\bar 1}),\quad i_1=1,i_2=3,i_{\bar 1}=2,i_{\bar 2}=4.
\]
We denote this D-soliton by $[1,2:3,4]$. Then identifying $[1,2:3,4]$ with the permutation
of the pairs, $(1,2)\leftrightarrow (3,4)$, one has the relation $[1,2:3,4]=[1:3]\cdot[2:4]=[1:4]\cdot[2:3]$
where $[i:j]$ is the permutation $(i)\leftrightarrow(j)$ of A-soliton. This relation
implies a resonant bifurcation of one D-soliton into two A-solitons (see Section \ref{twosoliton}).
Since one D-soliton has four $p_i$-parameters, we need to have $M=4N$ parameters
to describe $N$ D-soliton solution as a solution given by the $\tau$-function $\tau_N$.
However, we show in this paper that $4N$ parameter solution can contain up to $2N-1$ number of
D-solitons (this is quite different from the solitons of the KP equation).

In this paper, we study soliton solution of the DKP equation consisting of those
A- and D-type soliton solutions. In Section \ref{structure}, we give a general
structure of the $\tau$-functions. Then in Section \ref{onesoliton}, we discuss
some details of one D-soliton solutions given by the $B$-matrix of size $4\times 4$.
Here we also classify the soliton solutions obtained by $4\times 4$ $B$-matrices.
In Section \ref{twosoliton}, we present the soliton solutions for the case of
$8\times 8$ $B$-matrices, and classify the soliton solutions consisting of only
D-types. It turns out that the case with $8\times 8$ $B$-matrix can have either two or three
D-solitons depending on the values of the parameters $\{\,p_i\,:\,i=1\ldots 8\,\}$.
The generic solution given by the $B$-matrix having
all nonzero entries is then given by four A-solitons.
Finally we discuss the general case of $4N\times 4N$ $B$-matrices,
i.e. $M=4N$, in Section \ref{multisolitons}. Then we show that the number of D-solitons for
those $B$-matrices having $2N$ nonzero entries in the upper triangular part can be any
number from $N$ to $2N-1$, contrary to the previous study (see \cite{isojima:02}) where
the number of D-soliton is expected to be just $N$.

In \cite{kodama:04}, we showed that the $N$-soliton solutions of the KP equation can be classified
by the Schubert decomposition of the Grassmannian Gr$(N,2N)$.
We also expect that $N$-soliton solutions of the DKP equation can be classified by the similar
decomposition of the orthogonal Grassmannain OGr$(M,2M)$ for some $M$. However, in this paper, we
just present elementary feature of the $N$-soliton solutions based on the pfaffian structure
of the $\tau$ function, which is already complicated but has several interesting aspects.
We plan to discuss a classification problem based on the geometric structure of the
orthogonal Grassmannain in a future communication.


\section{Structure of the $\tau$-functions}\label{structure}
We first note that the $\tau$-function (\ref{tau}) with the ${\mathcal Q}_n$ matrix (\ref{Q})
has the following expansion theorem, i.e. the pfaffian version of the Binet-Cauchy theorem
(see \cite{ishikawa:95} for the details):
\begin{Lemma} \label{dtauBC} 
The $\tau$-functions of (\ref{tau}) with (\ref{Q}) can be expressed by
\[
\tau_n=\sum_{1\le i_1<\cdots<i_{2n}\le 4N} {\rm Pf}({B}(i_1,\ldots,i_{2n})){\rm Det}({E}(i_1,\ldots,i_{2n}))\,.
\]
where
${E}(i_1,\ldots,i_{2n})$ is $2n\times 2n$ submatrix of the $2n\times 4N$ matrix ${\mathcal E}_n$,
\[
E(i_1,\ldots,i_{2n}):=\begin{pmatrix}
E_{i_1} & \cdots & E_{i_{2n}} \\
\vdots  &  \ddots  & \vdots \\
E_{i_1}^{(2n-1)} & \cdots & E_{i_{2n}}^{(2n-1)}
\end{pmatrix}
\]
and ${B}(i_1,\ldots,i_{2n})$ is $2n\times 2n$ skewsymmetric submatrix of the $4N\times 4N$ matrix
${ B}$,
\[
{B}(i_1,\ldots,i_{2n})=\begin{pmatrix}
0 & b_{i_1,i_2} & \cdots & \cdots & b_{i_1,i_{2n}} \\
  &   0    & \ddots   &  \cdots &b_{i_2,i_{2n}}  \\
  &        &     \ddots            &  \ddots & \vdots \\
  &       &     & 0 & b_{i_{2n-1},i_{2n}}  \\
  &        &       &           &    0
  \end{pmatrix}
  \]
  The lower triangular part is given by $b_{i_l,i_k}=-b_{i_k,i_l}$, and is left in blank.
Also ${\rm Det}(E)$ is given by the Wronskian determinant,
\[
{\rm Det}(E(i_1,\cdots,i_{2n}))={\rm Wr}(E_{i_1},\ldots,E_{i_{2n}})=\prod_{m<n}(p_{i_n}-p_{i_m})\exp\left(\sum_{j=1}^{2n}\theta_{i_j}\right)\,>0.\]
The sign is due to the order (\ref{orderp}), i.e. $p_1<p_2<\cdots<p_{4N}$.
\end{Lemma}
 From this formula of the $\tau$ function, the soliton solutions are completely determined by the nonzero coefficients of the pfaffians Pf$(B(i_1,\ldots,i_{2n}))$ which are the Pl\"ucker coordinates of the orthogonal Grassmannan Gr$(2n,4N)$. For example, if all the coefficients are positive, then we have
 $(2n,4N-2n)$-soliton solution in $u(x,y,t)$, that is, we have $2n$ out-going line solitons in $y\to\infty$ and
 $4N-2n$ in-coming line solitons in $y\to-\infty$ (see \cite{biondini:03}). As we can see from
 \cite{kodama:04} that in particular, if $n=N$ we have an $2N$-soliton solution similar to the KP equation, i.e. all $2N$ solitons are A-solitons, and all
 the interactions are of T-types. Then the main purpose of this paper is to identify soliton solution
 of the DKP equation
 with a certain combination of nonzero coefficients.
 \begin{Example} As the simplest example,
the pfaffian Pf$({B}(i_1,\ldots,i_{4}))$ is given by 
\[
{\rm Pf}(B(1,\ldots,4))=b_{1,2}b_{3,4}-b_{1,3}b_{2,4}+b_{1,4}b_{2,3}\,.
\]
Note here that if the indices in each term have partial overlap, then the product
takes the minus sign, e.g. $(1,3)$ and $(2,4)$. This is a key for the classification
of four A-solitons (see Section \ref{twosoliton}).
\end{Example}

 From Lemma \ref{dtauBC}, one should note that each term in $\tau_n$ in (\ref{tau}) is given by
the product of $2n$ exponentials $E_{k}$, i.e.
\[
\prod_{j=1}^{2n}E_{i_j}=\prod_{k=1}^nE_{i_k,j_k}\,,
\]
where $\{\,i_j\,|\,j=1,\ldots,2n\,\}=\{\,i_k,j_k\,|\,
i_k<j_k,~k=1,\ldots,n\,\}$, and they are
the indices for the elements of the $B$-matrix, i.e. $b_{i_k,j_k}$. 
As we will show, 
this structure of the $\tau$-functions will be important to identify a soliton solution
in the asymptotics $y\to\pm\infty$.


\subsection{The $B$-matrix} 
Since the coefficients of the $\tau$-function are determined
by the $B$-matrix, we give some remarks on the structure of the matrix.
For a generic $4N\times 4N$ skew-symmetric matrix $B$, i.e. $B\in {\mathfrak{so}}(4N)$,
there is a decomposition called
{\it skew-Borel decomposition} \cite{adler:99},
\[
B=LJ_0L^T\,,
\]
where $L\in {\mathcal G}$, the group of invertible elements in the set of lower-triangular matrices
with nonzero $2\times 2$ blocks proportional to identity along the diagonal, i.e.
\[
{\mathcal G}:=\left\{\left.\begin{pmatrix}
a_1 & 0 & \cdots &\cdots & 0  &  0\\
0 & a_1 &\cdots  &\cdots & 0 & 0 \\
*  & *     &\ddots  &  \ddots & \vdots &\vdots\\
* & *     &  \ddots &\ddots  & \vdots  &\vdots\\
*  &   *  &  *  &* &    a_{2N}   &  0\\
*  &   *  &   * & * &  0  &  a_{2N}\\
\end{pmatrix}~\right|~\prod_{i=1}^{2N}a_i\ne 0~\right\}
\] 
The matrix $J_0$ is the $4N\times 4N$ skew-symmetric matrix whose $2\times 2$ diagonal
blocks are given by $\begin{pmatrix} 0& 1\\ -1&0\end{pmatrix}$ and all the other entries are zero,
i.e.
\begin{equation}\label{J0}
J_0:=\begin{pmatrix}
0 & 1 &  &  &    &  0 \\
-1 & 0 &   &  &   &   \\
   &       &\ddots  &  \ddots &   & \\
  &       &  \ddots &\ddots  &    & \\
   &      &    &  &    0   &  1\\
 0  &      &     &   &  -1  &  0
\end{pmatrix}
\end{equation}
A non-generic element of $\mathfrak{so}(4N)$ may be obtained by a permutation
matrix $\pi\in S_{4N}$ with $\pi^{-1}=\pi^T$ (i.e. $\pi\in O(4N)$), that is, we have
\[
B=\pi LJ_0L^T\pi^{-1}=(\pi L)J_0(\pi L)^T\,, 
\]
where $L$ is not a generic element in ${\mathcal G}$.
\begin{Example}\label{B4}
Consider the case $N=1$ ($4\times 4$ $B$-matrix). Then the generic element $B$ can be expressed by
$B=LJ_0L^T$ with
\[
L=\begin{pmatrix}
a & 0 & 0 & 0\\
0 & a & 0 & 0\\
b & d & f & 0\\
c&  e & 0 & f
\end{pmatrix}\, \in {\mathcal G}\,.
\]
Then the matrix $B$ is given by
\begin{equation}\label{genericB}
B=LJ_0L^T=\begin{pmatrix}
0 & a^2 & ad & ae \\
   & 0 & -ab & -ac \\
  &   &  0  & -cd + be+f^2\\
 &   &    &    0 
 \end{pmatrix}\,.
 \end{equation}
 (We leave the lower triangular part in blank for skew symmetric matrix.)
If a skew-symmetric matrix has zero at $(1,2)$-entry (i.e. $b_{1,2}=0$) and
others are nonzero, such matrix, say $B_1$, cannot be expressed
in this form. In this case, one can consider the form $B=(\pi L)J_0(\pi L)^T$ with
\[
w=s_{2,3}:=\begin{pmatrix}
1  &  0 & 0 & 0 \\
0 &  0  & 1 & 0\\
0  & 1  &  0  &  0\\
0 & 0 & 0 & 1
\end{pmatrix}\,, \quad  
L=\begin{pmatrix}
a & 0 & 0 & 0 \\
0 & a & 0 & 0\\
b & d & f & 0 \\
0 & e & 0 & f 
\end{pmatrix}\,.
\]
The matrix $B_1$ is then given by
\begin{equation}\label{B1}
B_1=(s_{2,3}L)J_0(s_{2,3}L)^{T}=\begin{pmatrix}
0  &  0  &  a^2  &  ae \\
 & 0 & ab & be+f^2\\
 & & 0 & -ad \\
 & & & 0 \\
 \end{pmatrix}\,.
 \end{equation}
 In particular, with $L=Id$, the $4\times 4$ identity matrix, we have $B_1=J_1$ defined by
 \begin{equation}\label{J1}
 J_1:=\begin{pmatrix}
 0 & 0 & 1 & 0 \\
  &  0  &  0  & 1\\
  &  &  0  & 0 \\
  & & & 0
  \end{pmatrix}\,.
  \end{equation}
Other non-generic element of $B$-matrix, say $B_2$, having $b_{1,2}=b_{1,3}=0$ can be
expressed by $B_2=(s_{2,4}L)J_0(s_{2,4}L)^{T}$, 
\begin{equation}\label{B2}
B_2=\begin{pmatrix}
0  &  0  &  0  & a^2 \\
   &  0   &  f^2  &  ac  \\
   &     &   0   &  -ab  \\
   &    &      &    0  \\
\end{pmatrix}\,,
\end{equation}
with
\[
s_{2,4}=\begin{pmatrix}
1  &  0  & 0 & 0 \\
0 &   0 & 0 & 1 \\
0 & 0 & -1 & 0 \\
0 & 1 & 0 & 0
\end{pmatrix}\, \quad {\rm and}\quad 
L=\begin{pmatrix}
a  &  0  &  0  &  0  \\
0  &  a  &  0  &  0  \\
b  &  0  &  f  &  0  \\
c  &  0  &  0  &  f 
\end{pmatrix}\,.
\]
In particular, for $L=Id$ we have $B_2=J_2$ defined by
\begin{equation}\label{J2}
 J_2:=\begin{pmatrix}
 0 & 0 & 0 &1 \\
  &  0  &  1  & 0\\
  &  &  0  & 0 \\
  & & & 0
  \end{pmatrix}\,.
\end{equation}
In the next section we show that those skew-symmetric matrices $J_0, J_1$ and $J_2$ 
define all the one D-soliton solutions.
\end{Example}

In general, one can start with $J_0$ to define the $\tau$-functions, that is,
$\tau_n={\rm Pf}({\mathcal Q_n})$ with ${\mathcal Q_n}={\mathcal E}_nJ_0{\mathcal E}_n^T$
of (\ref{Q}).
Then consider a generalization of the ${\mathcal Q}_n$-matrix by replacing $J_0$ with
$J_{\pi}:=\pi J_0 \pi^{-1} $ for some $\pi\in S_{4N}\cap O(4N)$. 
The role of $\pi$ can be understood as the change of order of the column vectors in ${\mathcal E}_n$,
i.e.
\[
{\mathcal E}_n\pi:=({\bf E}_1,{\bf E}_2,\ldots,{\bf E}_{4N})\pi=(\pm{\bf E}_{\pi(1)},\pm{\bf E}_{\pi(2)},\ldots,\pm{\bf E}_{\pi(4N)})\,,
\]
where the column vector ${\bf E}_k:=(E_k,E_k^{(1)},\ldots,E_k^{(2n-1)})^T$, and $\pi(k)$ indicates the permutation
$k\to \pi(k)$. The signs in $w$ should be chosen properly so that the new $\tau$-function
$\tau_n'=({\mathcal E}_n\pi)J_0({\mathcal E}_n\pi)^T$ is sign-definite, i.e.
non-singular (see Section \ref{twosoliton} for more details).


\section{One D-soliton and two A-solitons}\label{onesoliton}
Here we discuss the cases with $4\times 4$ skew-symmetric matrix $B$, and show that
the cases include one D-soliton and two A-solitons.
\subsection{D-solitons}
In Introduction, we show an example of one D-soliton (\ref{oneDsol}) whose $\tau$-function is given by
$\tau_1={\rm Pf}({\mathcal E}_1J_0{\mathcal E}_1^T)$ with $J_0$ in (\ref{J0}) and ${\mathcal E}_1$ in Lemma \ref{dtauBC} for $N=1$.
This D-soliton is denoted as $[1,2:3,4]$, since it exchanges the asymptotic values of
$w=(\ln\tau_1)_x$ as $(1,2) \leftrightarrow (3,4)$ (recall $(i,j):=p_i+p_j$).
A general form of the one D-soliton can be expressed as $[i_1,j_1:i_2,j_2]$ where
$i_1<i_2$ and $i_k<j_k$. There are three cases,
\[
[1,2:3,4],\quad [1,3:2,4],\quad [1,4:2,3]\,,
\]
which correspond to the D-solitons generated respectively by the skew-symmetric matrices,
$J_0, J_1$ and $J_2$ in Example \ref{B4} as the $B$-matrices. Recall here that
the soliton label $[i,j:k,l]$ indicates the nonzero elements $b_{i,j}$ and $b_{k,l}$ in
the $B$-matrix. Namely
for the $J_k$-matrix with nonzero entries with 1 at $(i_1,j_1)$ and $(i_2,j_2)$ in
the upper-triangular part, the $\tau$-function is given by
\begin{equation}\label{dtau1}
\tau_1=Q_{1,2}=(p_{j_1}-p_{i_1})E_{i_1}E_{j_1}+(p_{j_2}-p_{i_2})E_{i_2}E_{j_2}\,.
\end{equation}
Then the function $w=(\ln\tau_1)_x$ is
\begin{equation}\label{1Dsol}
w(x,y,t):=\frac{\partial}{\partial x}\ln\tau_1=\frac{1}{2}\sum_{k=1}^4p_{k}+\frac{1}{2}(p_{i_2}+p_{j_2}-p_{i_1}-p_{j_1})
{\rm tanh}\frac{1}{2}(\theta_{i_2,j_2}-\theta_{i_1,j_1})\,
\end{equation}
with $\theta_{i,j}=\theta_i+\theta_j+\ln|p_i-p_j|$.  
For each asymptotic of $x\to\pm\infty$, one of the exponential terms in the $\tau$-function (\ref{dtau1})
becomes dominant, and we have
\[
w(x,y,t)\to \left\{\begin{array}{llll}
p_{i_1}+p_{j_1},\quad {\rm as}~ x\to-\infty\,,\\
p_{i_2}+p_{j_2},\quad {\rm as}~ x\to\infty\,.
\end{array}
\right.
\]
Thus one D-soliton exchanges $(i_1,j_1) \leftrightarrow (i_2,j_2)$ with the values of $w$,
and is denoted by $[i_1,j_1:i_2,j_2]$, as an element of the Weyl group of D-type.

One soliton solution $u=2w_x$ is a plane wave having the form,
\[
u(x,y,t)=\phi(k_xx+k_yy-\omega t),
\]
Let us denote the wavenumber ${\bf k}=(k_x,k_y)$ and the frequency $\omega$ for (\ref{1Dsol})
by ${\bf k}[i_1,j_1:i_2,j_2]$ and $\omega[i_1,j_1:i_2,j_2]$, i.e.
\[\left\{
\begin{array}{lllll}
{\bf k}[i_1,j_1:i_2,j_2]=(p_{i_2}+p_{j_2}-p_{i_1}-p_{j_1},\, p_{i_2}^2+p_{j_2}^2-p_{i_1}^2-p_{j_1}^2),\\
{}\\
\omega[i_1,j_1:i_2,j_2]=-(p_{i_2}^3+p_{j_2}^3-p_{i_1}^3-p_{j_1}^3)
\end{array}\right.
\]
The slope (or velocity) of the soliton in the $x$-$y$ plane is
given by $c:=dx/dy=-k_y/k_x=-(p_{i_2}^2+p_{j_2}^2-p_{i_1}^2-p_{j_1}^2)/(p_{i_2}+p_{j_2}-p_{i_1}-p_{j_1})$. The peak of the soliton then determines the line in the $x$-$y$ plane given by the equation $\theta_{i_1,j_1}(x,y,t)=\theta_{i_2,j_2}(x,y,t)$ for each $t$.

\begin{figure}[t]
\includegraphics[width=7cm]{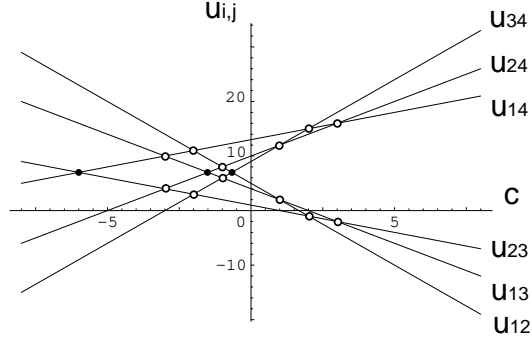}
\caption{The functions $u_{i,j}(c)$  and the velocities of solitons.
The intersection point between $u_{i,j}$ and $u_{k,l}$ gives the velocity
of D-soliton $[i,j:k,l]$. The solid dots give the velocities of D-solitons, and the open circles
give the velocities of A-solitons, e.g. the velocity of $[1:2]$-soliton is $c=-(p_1+p_2)$ which is
given by either $u_{1,4}=u_{2,4}$ or $u_{1,3}=u_{2,3}$.}
\label{uij:fig}
\end{figure}

We also note that the velocity $c$ of one D-soliton can be obtained as follows: 
First note that a D-soliton of $[i,j:k,l]$ gives a line $\theta_{i,j}=\theta_{k,l}$. Then setting $x=cy$ for
this equation, we have $\theta_{i,j}=u_{i,j}(c)y+\theta_{i,j}^0$ with 
\begin{equation}\label{uijc}
u_{i,j}(c):=(p_i+p_j)c+p_i^2+p_j^2\,.
\end{equation}
Then  taking the limit $|y|\to\infty$ for the equation $\theta_{i,j}=\theta_{k,l}$, 
the velocity $c$ of the D-soliton is determined by $u_{i,j}(c)=u_{k,l}(c)$, i.e. the intersection point
of $u_{i,j}$ and $u_{k,l}$. In Figure \ref{uij:fig}, we show $u_{i,j}(c)$ of (\ref{uijc}) for the
parameter ${\bf p}:=(p_1,\ldots,p_4)=(-2,-1,0,3)$.
Each solid dot gives the velocity of one D-soliton, and each open circle gives
the velocity of an A-soliton, e.g. the point at $u_{1,2}=u_{1,3}$ gives the velocity
of $[2:3]$-soliton. This implies that if $b_{1,2}$ and $b_{1,3}$ are only nonzero
elements in the $B$-matrix, then $\tau_1$ gives one A-soliton of $[2:3]$, i.e.
$\tau_1=(b_{1,2}(p_2-p_1)E_2+b_{1,3}(p_3-p_1)E_3)E_1$.
Figure \ref{uij:fig} can be also used to determine the solitons appearing
from the $\tau_1$-function in the asymptotics $|y|\to\infty$. For example,
if $b_{i,j}>0$ for all $1\le i<j\le 4$, then $\tau_1$ contains all six terms $E_iE_j$
with $i<j$.  Then from Figure \ref{uij:fig}, we have two A-soliton solutions of T-type,
$[1:3]$ and $[2:4]$ (see below for the details).

\subsection{Two A-solitons} One D-soliton can be considered as a {\it degenerate}
case of two A-solitons (recall that A-soliton exchanges single letter as $[i:j]$, while D-soliton
exchanges two letters as $[i,j:k,l]$). Namely, D-soliton of $[i,j:k,l]$ can be considered
as the ``product'' of two A-solitons of $[i:k]$ and $[j:l]$ or $[i:l]$ and $[j:k]$, i.e.
$[i,j:k,l]=[i:k]\cdot[j:l]=[i:l]\cdot[j:k]$ as the product of two permutations.
Then two A-soliton solutions can be obtained by adding extra exponential terms, i.e.
extra nonzero entries to the $B$-matrix. If one adds just one term, then we have 
Y-shaped solution satisfying the resonant condition. For example, for the $B$-matrix
having two nonzero entries $b_{i_1,j_1}$ and $b_{i_2,j_2}$ in the upper triangular part, if we add
a nonzero entry at either $(i_1,j_2)$ or $(i_2,j_2)$, we obtain a resonant Y-shaped soliton with
the resonant condition,
\[\left\{\begin{array}{cccc}
{\bf k}[i_1:i_2]+{\bf k}[j_1:j_2]&=&{\bf k}[i_1,j_1:i_2,j_2],\\
\omega[i_1:i_2]+\omega[j_1:j_2]&=& \omega[i_1,j_1:i_2,j_2]\,.
\end{array}\right.
\]
Here the wavenumber ${\bf k}[i:j]$ and the frequency $\omega[i:j]$ are for the one soliton solution of the KP equation, i.e.
\[
{\bf k}[i_1:i_2]=(p_{i_1}-p_{i_2}, p^2_{i_1}-p^2_{i_2}),\quad \omega[i_1:i_2]=p_{i_1}^3-p_{i_2}^3,
\]
\begin{figure}[t!]
\includegraphics[width=12cm]{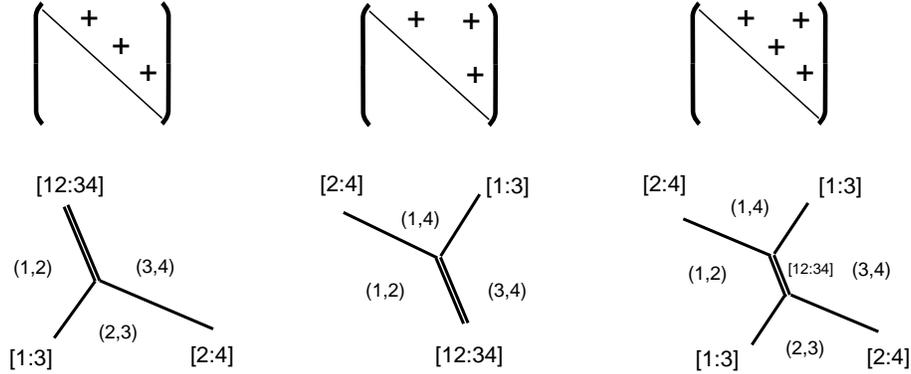}
\caption{Resonant bifurcations of the D-soliton $[1,2:3,4]$. Each region, $(i,j)$, corresponds to a dominant exponential marked by the element $b_{i,j}$. For example, with $b_{2,3}\ne 0$,
$[1,2:3,4]$-D-soliton bifurcates into $[1:3]$ and $[2:4]$ A-solitons as $y\to -\infty$ (top figure).
Those solitons can be found from Figure \ref{uij:fig}.}
\label{A4:fig}
\end{figure}
In Figure \ref{A4:fig} (see also Example \ref{1234Dsoliton} below), we show the resonant bifurcation of $[1,2:3,4]$-soliton into two
A-solitons of $[1:3]$ and $[2:4]$. Those solitons can be found from Figure \ref{uij:fig}.
For example, in the left figure of Figure \ref{A4:fig}, we have $u_{1,2}, u_{2,3}$ and $u_{3,4}$
from the nonzero entries of the $B$-matrix.
Then the three intersection points of the graphs of those $u_{i,j}$ in Figure \ref{uij:fig}
give one D-soliton (solid dot) and
two A-solitons (open circles). Figure \ref{uij:fig} also shows that the solid dot indicates the pair of
dominant exponentials $(E_{1,2},E_{3,4})$ for $[1,2:3,4]$-soliton as $y\to\infty$, while the open circles
show the dominant pairs $(E_{1,2}, E_{2,3})$ for $[1:3]$-soliton and $(E_{2,3},E_{3,4})$
for $[2:4]$-soliton as $y\to-\infty$.

Since there are at least four disconnected regions in the $x$-$y$ plane divided by two line solitons, one needs to add
at least two extra nonzero (positive) entries. Figure \ref{A4:fig} shows the resonant
{\it bifurcations} of a D-soliton into two A-solitons. Two soliton solution in this figure
is a new T-type (referred to as TD-type) which does not exist in the KP equation (i.e. $v^{\pm}\ne 0$). The T-type of the KP equation can be obtained by a $B$-matrix with Det$(B)=0$, e.g.
\[
B=\begin{pmatrix}
0 & 1 & 2 & 1\\
   & 0 & 1 & 1\\
   &   &  0  & 1\\
   &   &      & 0
   \end{pmatrix}
\]
Note $\tau_2=0$ with this matrix due to $Pf(B)=0$. One should also note that the generic skew-symmetric $B$-matrix having all nonzero entries gives a two A-soliton solution of T-type
(having a resonant hole), but it is not a solution 
of the KP equation. In this case, since those solitons are A-solitons, the function $v^+v^-=\tau_2/\tau_1^2$ appears
only locally in the $x$-$y$ plane.
Both $\tau$-functions for T-type two-soliton solutions of KP and DKP equations
have six independent exponentials, i.e. $E_{i,j}$ for $1\le i<j\le 4$, which is the key
to produce T-type resonant interaction of two A-solitons (see also \cite{biondini:03, kodama:04}).

\begin{figure}[t!]
\includegraphics[width=10.5cm]{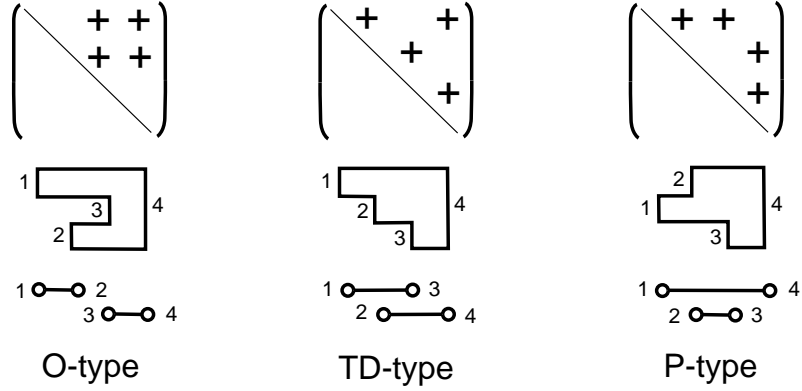}
\caption{Three fundamental types of two A-solitons obtained from D-soliton. Each diagram provides the relation between $B$-matrix and the corresponding two-soliton solution, i.e.
$b_{i,j}\ne 0$ if and only if the pair $(i,j)$ is directly connected in the middle diagram, and
$[k:l]$-soliton implies that the pair $(k,l)$ has no direct connection. For example, in O-type,
two A-solitons are labeled $[1:2]$ and $[3:4]$, then we have the middle diagram to
show the connections between the numbers in the pairs, i.e. (1,2) and (3,4) are not connected. This diagram implies we have nonzero elements $b_{1,3},b_{1,4},b_{2,3}$ and $b_{2,4}$.}
\label{A5:fig}
\end{figure}
\begin{figure}[t]
\includegraphics[width=11.5cm]{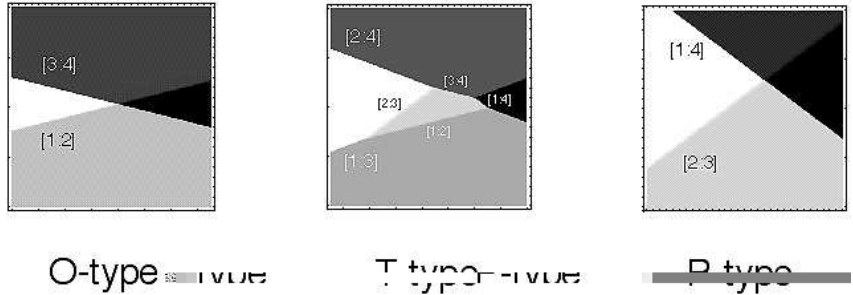}
\caption{Three fundamental types of two A-solitons.
Those solutions have the same topological feature as the two soliton solutions of the KP equation (see also Figure 3 in \cite{kodama:04}). The difference between those solutions appears as the existence
of localized solutions $v^{\pm}$ (if $v^{+}v^{-}=0$, then the DKP solution is also the KP solution).
However note that the TD-type in Figure \ref{A5:fig} does not have a resonant quadrangle.
The figures show the counter plots of the function $w=(\ln\tau)_x$ in the $x$-$y$ plane for
a fixed $t$, and the contrast of the shades shows the different values of $w$ which increases in the positive $x$-direction.}
\label{TPO:fig}
\end{figure}

\noindent
\begin{Example}\label{1234Dsoliton}
Consider $[1,2:3,4]$-D-soliton which corresponds to the $B$-matrix having nonzero entries at $(1,2)$ and $(3,4)$, i.e. $B=J_0$ in (\ref{J0}):
\begin{itemize}
\item[(a)]  $[1:3]$- and $[2:4]$-A-solitons (TD-type)  by putting $+1$ at the entries $(1,4)$ and $(2,3)$.
\item[(b)]  $[1:4]$- and $[2:3]$-A-solitons (P-type) by putting $+1$ at the entries $(1,3)$ and $(2,4)$.
\end{itemize}
Note that the TD-type in the case a) is {\it not} the same as the one of the KP-solitons,
i.e. there is no resonance in this case. Original T-type of two A-solitons
is obtained by the $B$-matrix having all
nonzero entries and Det$(B)=0$. Again there are three types of
two A-solitons for DKP.  These types play an important role for the classification problem
as a building block. Figure \ref{A5:fig} shows the diagrams of the three fundamental
types of two A-solitons. The middle diagrams of 8-gon connect the types with the corresponding
$B$-matrices as follows: We have nonzero $b_{i,j}$ if the pair $(i,j)$ is connected in the
diagram, and a corresponding A-soliton of $[k:l]$ is given by the pair $(k,l)$ having no connection
in the diagram. Also note that the corresponding $B$-matrix for O-type is not in the generic one
given in (\ref{genericB}), but in (\ref{B1}) (compare this with the O-type of KP solitons 
which belong to a Schubert cell of codimension one \cite{kodama:04}).
Figure \ref{TPO:fig} shows those types of two A-solitons, which are
the same as 2-soliton solutions of the KP equation (note here that we have set 
$\tau_2=0$ for all the types, i.e. $\tau_1$ gives a solution of the KP equation). The parameters are chosen as ${\bf p}=(-2,-1,0,3)$,
and the velocities of solitons can be found from Figure \ref{uij:fig}.
\end{Example}

\section{Solitons generated by $8\times 8$ $B$-matrices}\label{twosoliton}
Since each D-soliton has four phases, we need eight phases to describe
a two D-soliton solution. 
The $\tau_2$-function in the case with $8\times 8$ $B$-matrix is given by
\[
\tau_2=\sum_{1\le i_1<\cdots<i_4\le 8}{\rm Pf}({B}(i_1,\ldots,i_4))\Delta(i_1,\ldots,i_4)\exp\left(\sum_{k=1}^4
\theta_{i_k}\right)\,.
\]
where $\Delta(i_1,\ldots,i_4)=\prod_{1\le j<k\le4}(p_{i_j}-p_{i_k})>0$, and 
the pfaffian is calculated as
\[
{\rm Pf}({B}(i_1,\ldots,i_4))=b_{i_1,i_2}b_{i_3,i_4}-b_{i_1,i_3}b_{i_2,i_4}+b_{i_1,i_4}b_{i_2,i_3}\,.
\]

Let us now construct the $B$-matrices for two D-soliton solutions which
 consist of only four nonzero entries in the upper triangular parts.
We follow the steps:
 \begin{itemize}
\item[(1)] Take four pairs $(i_k,j_k),~k=1,..,4$ with $i_k<j_k$ from the set $\{1,\ldots,8\}$,  such that
\[
1=i_1<i_2<\cdots<i_4,\quad j_k\ne j_l,~~(k\ne l).
\]
There are $(2\times 4-1)!!=105$ ways to make those pairs (compare this with $({\bf n}^-,{\bf n}^+)$
in the case of 4 A-solitons \cite{kodama:04}).
Those four pairs give four nonzero elements in the $B$-matrix, i.e. $(i_k,j_k)\equiv b_{i_k,j_k}$
for $k=1,\ldots,4$.
\item[(2)] Determine the signs of each pairs $(i_k,j_k)$, so that all the pfaffians in the expansion of $\tau_2$ associated with the $B$-matrix have the same sign. Since the pairs $(i_k,j_k)$ are all different, each pfaffian
has just one term, denoted by ${\rm Pf}((i,j),(k,l))=\pm b_{i,j}b_{k,l}$. Here the sign depends on the
 pairs, for examples,
\[
{\rm Pf}((1,3),(5,7))=b_{1,3}b_{5,7},\quad {\rm Pf}((2,6),(4,8))=-b_{2,6}b_{4,8}.
\]
\end{itemize}
In order to determine the sign of the pfaffian ${\rm Pf}((i,j),(k,l))=\pm b_{i,j}b_{k,l}$, we define:
\begin{Definition} We say
that the pairs $(i_j,i_k)$ and $(i_l,i_m)$ with $i_j<i_l$ have a {\it partial} overlap, if $i_j<i_l<i_k<i_m$.
Otherwise, we say that the pairs have a {\it non}-partial overlap (i.e. total or no overlap). 
Let us introduce the sign of overlap between $(i_k,j_k)$ and $(i_l,j_l)$,
\[
\sigma_{kl}=\left\{\begin{array}{lllll}
- \quad {\rm if}\quad  i_k<i_l<j_k<j_l~~({\rm partial~overlap})\,,\\
+ \quad {\rm otherwise}\,.
\end{array}
\right.
\]
Then we have ${\rm Pf}((i_k,j_k),(i_l,j_l))=\sigma_{kl}b_{i_k,j_k}b_{i_l,j_l}$.
\end{Definition}

Now we have:
\begin{Lemma}
Suppose that the signs $\sigma_{kl}$ satisfiy
\[
\sigma_{12}\sigma_{13}\sigma_{14}=\sigma_{12}\sigma_{23}\sigma_{24}=\sigma_{13}\sigma_{23}\sigma_{34}=\sigma_{14}\sigma_{24}\sigma_{34}\,.
\]
Then one can make the $\tau$-function $\tau_2$ to be sign-definite, that is, all ${\rm Pf}({B}(i_1,\ldots,i_4))$ take the same sign.
\end{Lemma}
\begin{Proof}
We denote $\epsilon_k={\rm sgn}(i_k,j_k)$, so that sgn(Pf$((i_k,j_k)(i_l,j_l)))=\sigma_{kl}\epsilon_k\epsilon_l$.
The condition that all the terms have the same sign requires
\[
\sigma_{kl}\epsilon_k\epsilon_l=\sigma_{k'l'}\epsilon_{k'}\epsilon_{l'}\,,
\]
for any $k,l$ and $k',l'$. This leads to $\sigma_{12}\sigma_{34}=\sigma_{13}\sigma_{24}=\sigma_{14}\sigma_{23}$ which
gives the assertion.
\end{Proof}
Applying the Lemma, one can prove:

\begin{Proposition}\label{2Dsolitons}
Let ${\bf i}=(i_1,\ldots,i_4)$ and ${\bf j}=(j_1,\ldots,j_4)$.
Then there exist  33 cases of sign-definite $\tau$-functions whose $B$-matrices have nonzero entries only at $(i_k,j_k)$ 
and $(j_k,i_k)$ for $k=1,\ldots,4$.  Those cases are given by:
\begin{itemize}
\item[(a)] For ${\bf i}=(1,2,3,4)$, there are 8 cases with ${\bf j}=(j_1,\ldots,j_4)$;
\[
\begin{array}{llll}
(5,6,7,8),~(5,8,7,6),~(6,5,8,7),~(6,7,8,5),\\
(7,8,5,6),~(7,6,5,8),~(8,5,6,7),~(8,7,6,5)\,.
\end{array}
\]
\item[(b)] For ${\bf i}=(1,2,3,5)$, there are 4 cases,
\[
(4,7,6,8),~(4,7,8,6),~(6,7,4,8),~(8,7,4,6)\,.
\]
\item[(c)] For ${\bf i}=(1,2,3,6)$, there are 4 cases,
\[
(4,5,8,7),~(5,4,7,8),~(7,4,5,8),~(8,5,4,7)\,.
\]
\item[(d)] For ${\bf i}=(1,2,4,5)$, there are 4 cases,
\[
(3,8,6,7),~(3,6,8,7),~(6,3,7,8),~(8,3,7,6)\,.
\]
\item[(e)] For ${\bf i}=(1,2,4,6)$, there is one case,
$
(8,3,5,7)\,.
$
\item[(f)] For ${\bf i}=(1,2,3,7)$, there are two cases,
$ (4,5,6,8),~(6,5,4,8)\,.
$
\item[(g)] For ${\bf i}=(1,3,4,5)$, there are two cases,
$ (2,6,7,8),~(2,8,7,6)\,.
$
\item[(h)] For ${\bf i}=(1,2,4,7)$, there is one case,
$ (6,3,5,8)\,.
$
\item[(i)] For ${\bf i}=(1,3,4,6)$, there is one case,
$ (2,8,5,7)\,.$
\item[(j)] For ${\bf i}=(1,2,5,6)$, there are two cases,
$ (3,4,7,8),~(4,3,8,7)\,.
$
\item[(k)] For ${\bf i}=(1,2,5,7)$, there is one case,
$ (4,3,6,8)\,.$
\item[(l)] For ${\bf i}=(1,3,5,6)$, there is one case,
$ (2,4,8,7)\,.$
\item[(m)] For ${\bf i}=(1,3,4,7)$, there is one case,
$ (2,6,5,8)\,.$
\item[(n)] For ${\bf i}=(1,3,5,7)$, there is one case,
$ (2,4,6,8)\,.$
\end{itemize}
\end{Proposition}
The case (n) corresponds to the $B$-matrix $B=J_0$ defined in (\ref{J0}),
and the $B$-matrices for other cases can be expressed as $B=J_\pi:=\pi J_0\pi^{T}$ for
some $\pi\in S_{8}$ with appropriate signs in the entries of $\pi$. Those signs are
determined so that the $\tau$-function is non-singular.

\subsection{Two and three D-solitons}
We now classify D-solitons given by the $\tau$-functions associated with the $B$-matrices
obtained in Proposition \ref{2Dsolitons}.  We obtain:
\begin{Theorem}\label{2-3Dsolitons}
 Suppose we have the order, $(i_1,j_1)<(i_2,j_2)<(i_3,j_3)<(i_4,j_4)$ with
$(i_k,j_k):=p_{i_k}+p_{j_k}$.
Then as a generic situation, the D-soliton solutions of the DKP equation turns out to be the following
two cases;
\begin{itemize}
\item[(i)] two D-solitons of $[i_1,j_1:i_3,j_3], [i_2,j_2:i_4,j_4]$ or $[i_1,j_1:i_4,j_4], [i_2,j_2:i_3,j_3]$,
\item[(ii)] three D-solitons of $[i_k,j_k:i_{\alpha},j_{\alpha}], [i_l,j_l:i_{\alpha},j_{\alpha}]$ and
$[i_m,j_m:i_{\alpha},j_{\alpha}]$ for $\alpha=2$ or $3$ ($\alpha, k, l, m$ are all
distinct).
\end{itemize}
\end{Theorem}
\begin{Proof}
First we note that each $\tau_2$-function associated with the $B$-matrix given in Proposition \ref{2Dsolitons} has six exponential terms.
We denote them as $\hat E_k\hat E_l$ for $1\le k<l\le 4$, where
$\hat E_k:=E_{i_k,j_k}=\exp((p_{i_k}+p_{j_k})x+(p_{i_k}^2+p_{j_k}^2)y+\theta_{k}^0)$
for a fixed $t$. To identify two
dominant exponents which determine an asymptotic D-soliton,
we set $x=cy$ and define $u_{i,j}:=(p_{i}+p_{j})c+p_{i}^2+p_{j}^2$ (see Section \ref{onesoliton}).
Then for large $|y|$, the velocity $c$ of asymptotic soliton can be given by the intersection point
of $u_{i_k,j_k}=u_{i_l,j_l}$,
where as a generic situation only two exponential terms become
dominant, say $\hat E_k\hat E_m$ and $\hat E_l\hat E_m$ (see also \cite{biondini:03} for the details,
and Figure \ref{4768-sol:fig} below). The soliton is then identified as
$[i_k,j_k:i_l,j_l]$. Then there are two cases where the number of solitons, i.e. 
the number of balancing pairs of dominant exponentials, is either two
or three.
To show this, we consider, for example,  the case with ${\bf i}=(1,2,3,4),~{\bf j}=(5,6,7,8)$.
We set the parameter ${\bf p}$ so that $p^{(1)}_1<p^{(1)}_2<p^{(1)}_3<p^{(1)}_4$ whith
$p^{(1)}_k:=p_{i_k}+p_{j_k}$. Keeping this order, one can choose ${\bf p}$, so that
we have $p^{(2)}_4<p^{(2)}_3<p^{(2)}_2<p^{(2)}_1$ with $p^{(2)}_k:=p_{i_k}^2+p_{j_k}^2$.
(This can be done with a large negative value for $p_1$ and a small positive $p_8$.)
Then it is easy to see from the graph of $u_{k}:=u_{i_k,j_k}$ that one can have either
two D-soliton of $[i_1,j_1:i_3,j_3]$ and $[i_2,j_2:i_4,j_4]$ or three D-soliton of
$[i_1,j_1:i_3,j_3], [i_2,j_2:i_3,j_3]$ and $[i_3,j_3:i_4,j_4]$. The parameters ${\bf p}=(p_1,\ldots,p_8)$
can take the same values except $p_8$ for both cases (Figure \ref{4768-sol:fig}
gives an example for this, also see below).

One can also make a different order for $p^{(2)}_k$, e.g.
$p^{(2)}_3<p^{(2)}_2<p^{(2)}_1<p^{(2)}_4$ (by taking large positive $p_8$).
Then one can have three D-soliton of $[i_1,j_1:i_2,j_2],[i_2,j_2:i_3,j_3]$ and $[i_2,j_2:i_4,j_4]$
with the velocities $c_{2,4}<c_{1,2}<c_{2,3}$ where $c_{k,l}$ is the velocity of
$[i_k,j_k:i_l,j_l]$-soliton (recall that those are given by
the intersection $u_{k}=u_{l}$ with $u_k=u_{i_k,j_k}$).

Thus there is a freedom in the ordering of $p^{(2)}_k$, and this is a key to
generate three D-solitons (compare with the case for the KP equation \cite{biondini:03, kodama:04}
where we have only two A-solitons for six exponential terms in the $\tau$-function).

 In Figure \ref{4768-sol:fig}, we illustrate the case of ${\bf i}=(1,2,3,5)$ and ${\bf j}=(4,7,6,8)$:
Here the ordering is $(1,4)<(2,7)<(3,6)<(5,8)$. Then depending on the parameters $p_k$'s,
the number of asymptotic D-solitons is either two or three.
The left (right) figure shows the case of two (three) D-solitons, i.e. there are two (three)
values of $c$ where two exponential terms become dominant. It is obvious that there is
no case other than those two in the general case.
\end{Proof}

\begin{figure}[t]
\includegraphics[width=10cm]{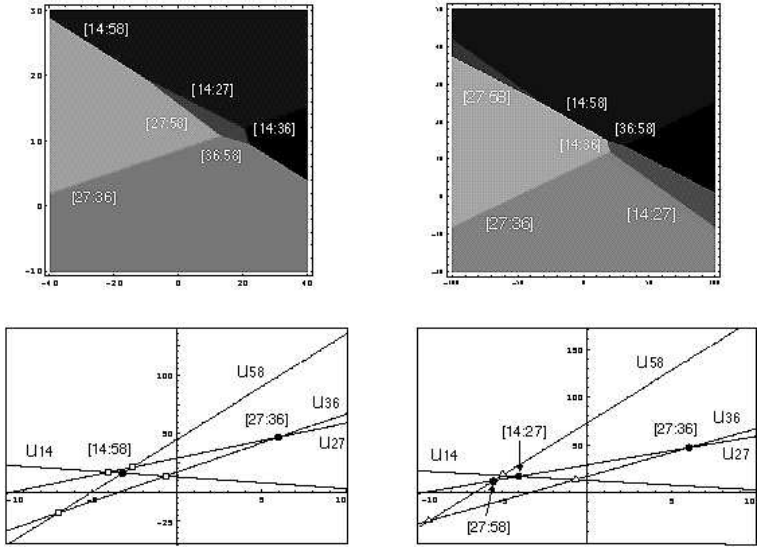}
\caption{Two and three D-solitons associated with ${\bf i}=(1,2,3,5)$ and ${\bf j}=(4,7,6,8)$.
The left figure shows two D-solitons with ${\bf p}=(-3,-2,1,2,3,4,5,6)$. The right one shows three D-solitons with the same parameters except $p_8=8$. The bottom figures show the graphs of $u_{i,j}(c)=(p_i+p_j)c+p^2_i+p^2_j$. The dots in the graphs show the dominant two exponential terms, which determine the asymptotic D-solitons. The squares
correspond to the intermediate D-solitons forming the resonant quadrangle, and the triangles correspond to
the resonant intermediate solitons from the pairs of three asymptotic solitons as shown in the top figures.}
\label{4768-sol:fig}
\end{figure}

We note here that all cases in Proposition \ref{2Dsolitons} have two-D-soliton solutions, but only
some cases can have three-D-soliton solution. In particular, we have:
\begin{Proposition}\label{case-n}
The case (n) in Proposition \ref{2Dsolitons} can have only two-D-soliton solutions
of $[1,2:5,6]$ and $[3,4:7,8]$.
\end{Proposition}
In order to prove the Proposition, we prepare the following Lemma:
Let us first denote $u_{i_k,j_k}$ as $u_k$ as before:
\[
u_k(c):=(p_{i_k}+p_{j_k})c+p_{i_k}^2+p_{j_k}^2\,,\quad k=1,\ldots,4\,,
\]
where $(i_k,j_k)=(2k-1,2k)$ for the $B$-matrix of the case (n), i.e. $B=J_0$.
Then we have:
\begin{Lemma}\label{cij}
Let $c_{j,k}$ be the point of intersection given by $u_j(c)=u_k(c)$.
Then for a fixed index $\alpha\in\{1,\ldots,4\}$, we have
\[
c_{\alpha, k}<c_{\alpha, j}<c_{\alpha, i}\,,\quad {\rm for}\quad i<j<k\,,~~\alpha\ne i,j,k\,.
\]
\end{Lemma}
\begin{Proof}
 From $u_k(c)=u_j(c)$, we have
 \[
 c_{j,k}=-\frac{(p_{2j-1}^2+p_{2j}^2)-(p_{2k-1}^2+p_{2k}^2)}{(p_{2j-1}+p_{2j})-(p_{2k-1}+p_{2k})}\,.
 \]
 Let us write $p_{2k-1}$ and $p_{2k}$ as
 \[
 p_{2k-1}=P_k-\Delta_k,\quad p_{2k}=P_k+\Delta_k\,,
 \]
that is, $P_k=\frac{1}{2}(p_{2k-1}+p_{2k})$, and we have
\[
c_{j,k}=-(P_j+P_k)-\frac{\Delta_j^2-\Delta_k^2}{P_j-P_k}\,.
\]
 Suppose $\alpha<j<k$ (other cases follow the similar argument). Then we have
 \[
 c_{\alpha, j}-c_{\alpha, k}=P_k-P_j+\frac{\Delta_{\alpha}^2-\Delta_k^2}{P_{\alpha}-P_k}
 -\frac{\Delta_{\alpha}^2-\Delta_j^2}{P_{\alpha}-P_j}\,.
 \]
It is then easy to show $c_{\alpha, j}-c_{\alpha, k}>0$. 
 \end{Proof}
 \begin{figure}[t]
\includegraphics[width=6cm]{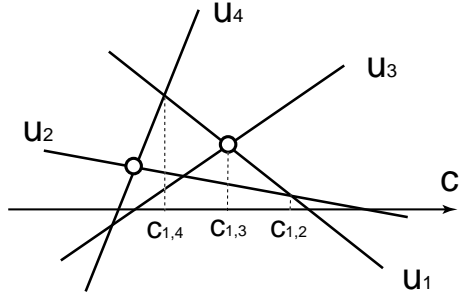}
\caption{The qualitative graphs of the functions $u_k(c)=(p_{2k-1}+p_{2k})c+p_{2k-1}^2+p_{2k}^2$ for the case (n) in Proposition
\ref{2Dsolitons} which states, e.g. $c_{1,4}<c_{1,3}<c_{1,2}$. The open circles correspond to two D-solitons of
$[1,2:5,6]$ with the velocity $c_{1,3}$ and $[3,4:7,8]$ with $c_{2,4}$.}
\label{c:fig}
\end{figure}

 \medskip
 \noindent
 Now we give a proof of Proposition \ref{case-n}:
 \begin{Proof}
  From Lemma \ref{cij}, we have the order $c_{1,2}<c_{1,3}<c_{1,4}$. Also note that the slope of $u_k(c)$ is given by $p_{2k-1}+p_{2k}=:p^{(1)}_k$, hence $p^{(1)}_1<p^{(1)}_2<p^{(1)}_3<p^{(1)}_4$. Then we obtain qualitative
graphs of $u_k$ as shown in Figure \ref{c:fig}, and from the graphs, we can find the dominant pairs of the exponential terms in $\tau_2$ for $y\to\pm\infty$. This shows that the dominant pairs are given by
$(\hat E_1\hat E_4, \hat E_3\hat E_4)$ and $(\hat E_1\hat E_4,\hat E_1\hat E_2)$ for $y\to\infty$,
and $(\hat E_1\hat E_2,\hat E_2\hat E_3)$ and $(\hat E_2\hat E_3,\hat E_3\hat E_4)$ for
$y\to-\infty$.  This implies that we have two D-solitons of $[1,2:5,6]$ and $[3,4:7,8]$
(recall $\hat E_k=E_{2k-1,2k}$). This completes the proof.
\end{Proof}

\begin{figure}[t]
\includegraphics[width=11.5cm]{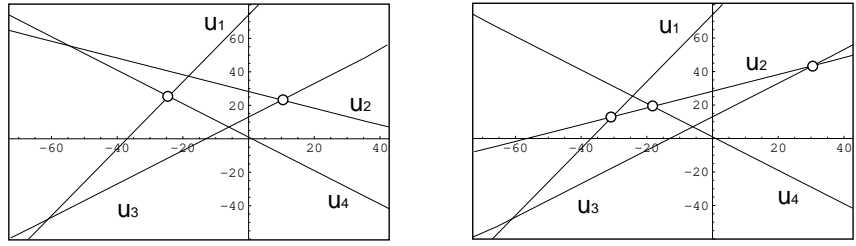}
\caption{The graphs of $u_k(c)=(p_{k}+p_{9-k})c+p_{k}^2+p_{9-k}^2$ for $k=1,\ldots,4$.
The indices are marked by $(i_k,j_k)=(k,9-k)$ for $k=1,\ldots,4$.
The left figure shows two D-soliton solutions of $[1,8:4,5]$ and $[2,7:3,6]$ with 
${\bf p}=(-5,-4,-2,-1,0,3,3.5,7)$, and the right one shows three D-solitons of $[1,8:2,7],[2,7:3,6]$ and $[2,7:4,5]$ with
${\bf p}=(-5,-3.5,-2,-1,0,3,4,7)$. Note that only $p_2$ is changed.}
\label{case-a:fig}
\end{figure}

\begin{figure}[t]
\includegraphics[width=11.5cm]{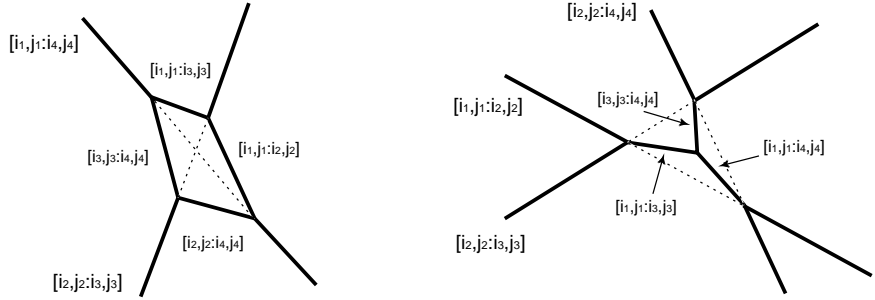}
\caption{Interaction patterns for two and three D-solitons in Figure \ref{case-a:fig}.
The function $w=(\ln\tau_2)_x$ takes the values, $w\to (i_1,j_1,i_2,j_2)$
as $x\to-\infty$, and $w\to(i_3,j_3,i_4,j_4)$ as $x\to\infty$. All six D-solitons
in each figure can be identified as the six intersection points of the graphs $u_k:=u_{i_k,j_k}$ in Figure \ref{case-a:fig}.}
\label{3Dsol:fig}
\end{figure}

We also have:
\begin{Proposition}\label{case-a}
The case with ${\bf i}=(1,2,3,4)$ and ${\bf j}=(8,7,6,5)$ (one of the cases in (a) of Proposition \ref{2Dsolitons}) can have both
two and three D-solitons.
\end{Proposition}
\begin{Proof} Follow the argument in the proof of Theorem \ref{2-3Dsolitons}.
Also see the following examples:
\begin{itemize}
\item[(i)] For two D-solitons, we take 
${\bf p}:=(p_1,p_2,\ldots,p_8)=(-5,-4,-2,-1,0,3,3.5,7)$.
Then the open circles in the left figure in Figure \ref{case-a:fig} show the dominant exponentials
which correspond to the D-solitons of $[i_1,j_1:i_4,j_4]=[1,8:4,5]$ and $[i_2,j_2:i_3,j_3]=[2,7:3,6]$.
\item[(ii)] For three D-solitons, we take ${\bf p}=(-5,-3.5,-2,-1,0,3,4,7)$. The three open circles in the right figure in Figure \ref{case-a:fig} correspond to the two D-solitons of $[i_1,j_1:i_2,j_2]=[1,8:2,7]$,
$[i_2,j_2:i_3,j_3]=[2,7:3,6]$ and $[i_2,j_2:i_4,j_4]=[2,7:4,5]$. 
\end{itemize}
In Figure \ref{3Dsol:fig}, we show the interaction patterns for those cases shown in Figure \ref{case-a:fig}
(they are topologically the same as in Figure \ref{4768-sol:fig}).
\end{Proof}

In Section \ref{multisolitons}, we will extend Propositions \ref{case-n} and \ref{case-a} to the general cases as Propositions \ref{multi-n} and \ref{multi-a}.

Recall that a D-soliton can be identified as an element of the Weyl group of D-type, $W^D$. In the case of two D-solitons, we have, for example,
$\pi_{1,3}: (i_1,j_1)\leftrightarrow (i_3,j_3)$ and $\pi_{2,4}: (i_2,j_2)\leftrightarrow (i_4,j_4)$.
The group generated by those elements is an abelian subgroup of $W^D$, and
the orbit of this subgroup of $(i_1,j_1,i_2,j_2)$ represents the asymptotic regions 
divided by the solitons, i.e. $(i_1,j_1,i_2,j_2), (i_1,j_1,i_4,j_4),(i_2,j_2,i_3,j_3)$ and $(i_3,j_3,i_4,j_4)$,
which correspond to the dominant exponentials in the $\tau$-function, that is,
$\hat E_1\hat E_2, \hat E_1\hat E_4, \hat E_2\hat E_3$ and $\hat E_3\hat E_4$, respectively.
In the $\tau$-function, we have two more exponential terms labeled by $(i_1,j_1,i_3,j_3)$ and
$(i_2,j_2,i_4,j_4)$, i.e. $\hat E_1\hat E_3$ and $\hat E_2\hat E_4$. Those terms become dominant at the points of the resonant quadrangle 
as in the case of two A-solitons of T-type. Namely we have a resonant two D-soliton.
The resonant condition is given by, for example,
\[\left\{
\begin{array}{llll}
{\bf k}[i_1,j_1:i_3,j_3]+{\bf k}[i_3,j_3:i_4,j_4]={\bf k}[i_1,j_1:i_4,j_4]\\
{}\\
\omega[i_1,j_1:i_3,j_3]+\omega[i_3,j_3:i_4,j_4]=\omega[i_1,j_1:i_4,j_4]
\end{array}\right.
\]
In the case of three D-soliton, for example $[i_1,j_1:i_2,j_2],[i_2,j_2:i_3,j_3]$ and $[i_2,j_2:i_4,j_4]$,
we have $\pi_{1,2}, \pi_{2,3}$ and $\pi_{2,4}$ as the generating elements of the subgroup.
Then the subgroup contains $\pi_{1,3}, \pi_{1,4}$ and $\pi_{3,4}$, and those are the 
D-solitons generated by the resonances of pairs, i.e.
$\pi_{k,l}$ represents D-soliton of $[i_k,j_k:i_l,j_l]$.  Each resonant relation can be represented by
the product $\pi_{i,2}\cdot \pi_{2,j}=\pi_{i,j}$.

For any two D-soliton solution in Proposition \ref{2Dsolitons}, we have:
\begin{Proposition}
Two D-solitons constructed above are all in resonance, in the sense that there is a resonant quadrangle at the 
intersection point. Namely all the cases are of T-type with one hole.
\end{Proposition}
\begin{Proof}
First note that there are six exponential terms in the $\tau$-function.
Each exponential term becomes dominant in some region in the $x$-$y$ plane.
For example, consider two D-solitons with $[1,2:5,6]$ and $[3,4:7,8]$.
Then the function $w=(\ln \tau_2)_x$ takes $w\to (1,2,3,4)$ for $x\to-\infty$, and $w\to (5,6,7,8)$
for $x\to\infty$.
With those D-solitons, we have the regions marked by $(1,2,7,8)$ and $(3,4,5,6)$.
In addition to those exponentials, we also have $(1,2,5,6)$ and $(3,4,7,8)$.
As in the case of two solitons of the KP equation (see \cite{biondini:03}),
a resonant interaction of Y-shape can occur. For example,
$[1,2:5,6], [1,2:7,8]$ and $[5,6:7,8]$ solitons are in resonant,
and the three regions divided by those solitons are marked by
$(1,2,3,4), (3,4,5,6)$ and $(3,4,7,8)$. Then the six exponentials
in the $\tau$-function form two D-solitons in resonance (see the left figure in Figure \ref{3Dsol:fig}). 
\end{Proof}

\begin{Remark}
In Figure 8 of \cite{isojima:02}, Isojima et al presented an example of (two) D-solitons
which seems to have only one intermediate soliton (no resonant quadrangle). 
However, it turns out that this is {\it not} two D-solitons:
First we note that this
 is the case (k) in Proposition \ref{2Dsolitons}, i.e.
${\bf i}=(1,2,5,7)$ and ${\bf j}=(4,3,6,8)$. This can be found by ordering
the parameters $(p_1,\ldots,p_4,q_1,\ldots,q_4)$ in Figure 8 of \cite{isojima:02},
i.e. relabel $(p_1,\ldots,q_4)$ in terms of our ordering $(p_1,\ldots,p_8)$
with $p_1\to p_8, p_2\to p_7,p_3\to p_4,p_4\to p_1,q_1\to p_6,q_2\to p_5,q_3\to p_3$ and $q_4\to p_2$.
The parameters chosen in the paper is then given by ${\bf p}=(-\frac{3}{2},-\frac{4}{3},-\frac{2}{3},-\frac{1}{2},\frac{1}{2},\frac{2}{3},\frac{3}{2},3)$.
 With those parameters, we note that $p_1+p_4=:(1,4)=(2,3)=:p_2+p_3$, a {\it degeneracy} in the parameters.
Two D-solitons are given by $[1,4:5,6]$ and $[2,3:7,8]$, and the asymptotic
values of $w$ are $(1,2,3,4)$ and $(5,6,7,8)$ for $x\to\mp\infty$. 
Then the resonant quadrangle consists of the D-solitons of $[1,4:2,3], [1,4:7,8], [2,3:5,6]$ and
$[5,6:7,8]$. However, because of the degeneracy $(1,4)=(2,3)$, the $[1,4:2,3]$-soliton cannot exist,
i.e. the function $w$ cannot change the value across this soliton.
Also notice that the $[2,3:5,6]$- and $[1,4:5,6]$-solitons are almost parallel, and so are
the $[1,4:7,8]$- and $[2,3:7,8]$-solitons: For example, the peak of the soliton $[2,3:5,6]$ is given by
$\theta_{2,3}=\theta_{5,6}$, i.e. for a fixed $t$ and with the values of ${\bf p}$, we have
the line for $[2,3:5,6]$-soliton,
\[
(p_5+p_6-p_2-p_3)x+(p_5^2+p_6^2-p_2^2-p_3^2)y=\frac{57}{18}\left(x-\frac{55}{114}y\right)={\rm const.}
\]
which is almost parallel to the line $\theta_{1,4}=\theta_{5,6}$ for the $[1,4:5,6]$-soliton,
\[
(p_5+p_6-p_1-p_4)x+(p_5^2+p_6^2-p_1^2-p_4^2)y=\frac{57}{18}\left(x-\frac{65}{114}y\right)={\rm const.}
\]
Thus the soliton solution in Figure 8 of
\cite{isojima:02} consists of two D-solitons of $[1,4:5,6]$ and $[2,3:7,8]$ for $y\to\infty$
and two {\it other} D-solitons of $[2,3:5,6]$ and $[1,4:7,8]$ for $y\to -\infty$, and the intermediate
soliton is the $[5,6:7,8]$-soliton. Other intermediate soliton of $[1,4:2,3]$ may be considered to be
located at $y=-\infty$.
\end{Remark}

For any three D-soliton solutions in Proposition \ref{2Dsolitons}, we have:
\begin{Proposition}
Any pair of three D-solitons constructed above is in resonance to form
an Y-shape vertex. This three D-soliton solution has no resonant hole and 
four resonant Y-shape vertices.
\end{Proposition}
\begin{Proof}
Since there are six exponential terms which mark the six
separated regions in the $x$-$y$ plane bounded by those three solitons,
this 3-soliton solution cannot have a resonant hole. 
Three solitons are labeled by, for example, $[i_1,j_1:i_2,j_2], [i_2,j_2:i_3,j_3]$ and $[i_2,j_2:i_4,j_4]$,
and any pair of those solitons is in resonance, i.e. $[i_1,j_1:i_2,j_2]$ and $[i_2,j_2:i_3,j_3]$ have
a resonant interaction to generate $[i_1,j_1;i_3,j_3]$-soliton (see the right figure in Fig.\ref{3Dsol:fig}).
It is easy to see that this resonance appear for any three solitons obtained in Proposition \ref{2Dsolitons}.
\end{Proof}


\subsection{Three soliton solutions consisting of two A-solitons and one D-soliton}
Since one D-soliton is a degenerate two A-solitons, one can construct 
3-soliton solution consisting of two A-solitons and one D-soliton from two D-solitons. 
One should note that resolution of one D-soliton into two A-solitons can work 
only for the case of two D-solitons. In the case of three D-solitons, the resolution
of one D-soliton affects to other D-solitons, and it cannot produce two A-solitons.
This can be checked by the asymptotic analysis using the graphs of $u_{i,j}$.

\begin{figure}[t]
\includegraphics[width=9cm]{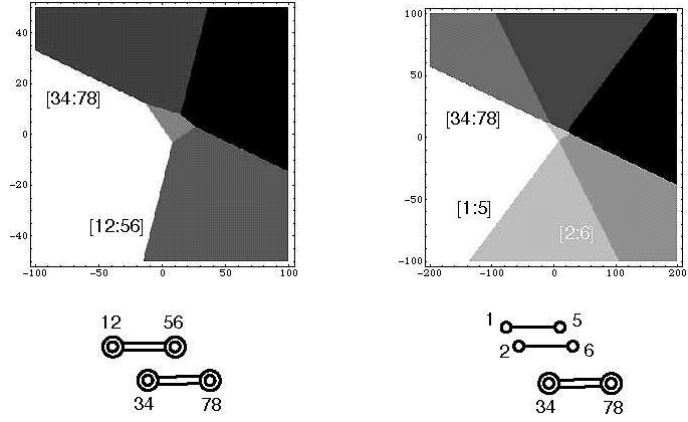}
\caption{Bifurcation of two D-solitons into three solitons with two A-solitons and one D-soliton.
The $[1,2:5,6]$-D-soliton resolves into two A-solitons of $[1:5]$ and $[2:6]$.
D-solitons are expressed as the double lines joined with the double circles.}
\label{A7:fig}
\end{figure}

\begin{Example}
Consider two D-solitons
corresponding to the $B$-matrix of the case (n) in Proposition \ref{2Dsolitons}, i.e.
${\bf i}=(1,3,5,7)$ and ${\bf j}=(2,4,6,8)$.
Two D-solitons are $[1,2:5,6]$ and $[3,4:7,8]$. Then, for example, 
two A-soliton of type $[1:5]$ and $[2:6]$ are obtained by adding two extra nonzero elements 
at $(1,6)$ and $(2,5)$. This fact can be verified easily from the diagram
of T-type (up-side-down staircase with sides labeled by 1,2,5 and 6; see Figure \ref{A5:fig}). The 3-solitons obtained here
has only one hole (see Figure \ref{A7:fig}).

One can also get two A-solitons of $[1:6]$ and $[2:5]$
from $[1,2:5,6]$-soliton by adding two nonzero elements at $(1,5)$ and $(2,6)$. Notice that
those two A-solitons form the diagram of P-type (cannon shape with sides by 1,2,5 and 6;
see Figure \ref{A5:fig}).

In the same way, one can resolve $[3,4:7,8]$-solitons into either the pair of $[3:7]$ and $[4:8]$ or 
the pair of $[3:8]$ and $[4:7]$, by adding extra nonzero entries to the $B$-matrix of the case (n).
\end{Example}

\subsection{Four A-soliton solutions from D-solitons}

Those solitons can be constructed in the similar way as the case of three solitons (two A-solitons
and one D-soliton)
from two D-solitons. However,
the compatibility among those A-solitons separated from D-solitons is somewhat complicated.  
For example, one can expect that two D-solitons, $[1,2:5,6]$ and $[3,4:7,8]$,
generate two A-solitons $[1:5], [2:6]$ from $[1,2:5,6]$ and other two A-solitons $[3:7]$ and $[4:8]$
from $[3,4:7,8]$. 
The $B$-matrices for 3-solitons $[1:5], [2:6], [3,4:7,8]$ and $[1,2:5,6], [3:7], [4:8]$ may be respectively given by
\[
\begin{pmatrix}
0 & 1 & 0 & 0 & 0 & 1 & 0 &0\\
  & 0 & 0 & 0 & 1 & 0 & 0 &0\\
  &   & 0 & 1 & 0 & 0 & 0 & 0\\
  &   &   & 0 & 0 & 0 & 0 & 0\\
  &   &  &   & 0 & 1 & 0 & 0\\
  &   &   &   &   & 0 & 0 & 0\\
  &   &   &   &   &   & 0 & 1\\
  &  &   &   &   &   &  & 0\\
\end{pmatrix} \quad{\rm and} \quad
\begin{pmatrix}
0 & 1 & 0 & 0 & 0 & 0 & 0 &0\\
  & 0 & 0 & 0 & 0 & 0 & 0 &0\\
  &   & 0 & 1 & 0 & 0 & 0 & 1\\
  &   &   & 0 & 0 & 0 & 1 & 0\\
  &   &  &   & 0 & 1 & 0 & 0\\
  &   &   &   &   & 0 & 0 & 0\\
   &   &   &   &   &   & 0 & 1\\
      &  &   &   &   &   &  & 0\\

\end{pmatrix}\,.
\]
Then adding $b_{3,8}, b_{4,7}$ to the $B$-matrix in the left (or $b_{1,6}, b_{2,5}$ to
the right one), the resulting $B$-matrix is expected to give four A-solitons $[1:5],[2:6],[3:7]$ and $[4:8]$.
However this matrix gives a singular solution, and a correct $B$-matrix for this 4-soliton solution
may be given by
\[
\begin{pmatrix}
0 & 1 & 0 & 0 & 0 & 0 & 0 &1\\
  & 0 & 1 & 0 & 0 & 0 & 0 &0\\
  &   & 0 & 1 & 0 & 0 & 0 & 0\\
  &   &   & 0 & 1 & 0 & 0 & 0\\
  &   &  &   & 0 & 1 & 0 & 0\\
  &   &   &   &   & 0 & 1 & 0\\
   &   &   &   &   &   & 0 & 1\\
      &  &   &   &   &   &  & 0\\
\end{pmatrix}\,.
\]
Thus in this case we do not have a compatibility between the two sets of 3 solitons obtained from the D-solitons (in the sense that the $\tau$-function cannot stay to be sign-definite).
 A general rule to construct a $B$-matrix for
 four A-solitons from two D-solitons seems to be complicated.  Here 
we give  a direct way to construct a $B$-matrix for a given four A-solitons,
based on the $B$-matrices obtained in Proposition \ref{2Dsolitons}:
\begin{itemize}
\item[1)] Give four A-solitons labeled by a pair of four numbers, i.e.
\[
{\bf n}^-=(n_1^-,\ldots,n_4^-),\quad  {\bf n}^+=(n_1^+,\ldots,n_4^+)
\]
where $1=n_1^-<\cdots<n_4^-$ and $n_i^-<n_i^+$. Then each soliton is labeled by
$[n_k^-:n_k^+]$ for $k=1,\ldots,4$ (the same as in the KP solitons \cite{kodama:04}).
This implies that the asymptotic values of $w=( \ln\tau)_x$ are given by
$\sum_{k=1}^4 p_{n_k^{\pm}}$ for $x\to\pm\infty$.
\item[2)]  Take a pair of two solitons, say $[n_i^-:n_i^+]$ and $[n_j^-:n_j^+]$, and
draw the corresponding diagrams shown in Figure \ref{A5:fig}, and do the same for the other pair.
Here the choice of the pair should be consistent with the asymptotic values.
Then adjust the signs for each horizontal edges of the diagrams, so that all
the pfaffians have the same sign. If all are satisfied, then the diagrams give
the $B$-matrix.
\end{itemize}
\noindent
{\bf Example:}
Consider four solitons with $[1:2], [7:8]$ and any other two solitons with the indices from $\{3,4,5,6\}$.   
Then one can construct the $B$-matrix as follows:

The diagram corresponding to $[1:2],[7:8]$ is a backward C-shape (P-type shown in Figure \ref{A5:fig}), having
1,2,7 and 8 on the vertical edge. This implies we have nonzero elements for $b_{1,7},b_{1,8},
b_{2,7},b_{2,8}$. Now if the other pair is $[3:6],[4:5]$, then the diagram associated to this is
a cannon-shape (O-type) having 3,4,5,6 on the vertical edges. This gives the nonzero entries
$b_{3,4},b_{3,5},b_{4,6},b_{5,6}$. In this example, all those entries in the upper triangular
part can take $+1$ for a sign-definite $\tau$-function.  Namely we have
\begin{equation}\label{1235B}
B=\begin{pmatrix}
0 & 0 & 0 & 0 & 0 & 0 & 1 &1\\
  & 0 & 0 & 0 & 0 & 0 & 1 &1\\
  &   & 0 & 1 & 1 & 0 & 0 & 0\\
  &   &   & 0 & 0 & 1 & 0 & 0\\
  &   &   &   & 0 & 1 & 0 & 0\\
  &   &   &   &   & 0 & 0 & 0\\
  &   &   &   &   &   & 0 & 0\\
  &   &   &   &   &   &  & 0\\
\end{pmatrix}
\end{equation}
In the case of two solitons of T-type, $[3:5],[4:6]$, one can use the $B$-matrix 
having the same entries except $b_{3,5}=2$, and in addition $b_{3,6}=1,b_{4,5}=1$.

Notice that the nonzero entries has two groups corresponding to two D-solitons $[1,7:2,8]$ and $[3,4:5,6]$, and
those groups have no partial overlaps. This structure is true for any cases satisfying the nonsigular condition, i.e. all the pfaffians have the same sign. (Recall that the $B$-matrix with $b_{1,8}=b_{2,7}= b_{3,4}=b_{5,6}=1$, i.e. ${\bf i}=(1,2,3,5), {\bf j}=(8,7,4,6)$, also gives three D-solitons of $[1,8:2,7],[2,7:3,4]$ and $[2,7:5,6]$.)

\begin{figure}[t]
\includegraphics[width=13cm]{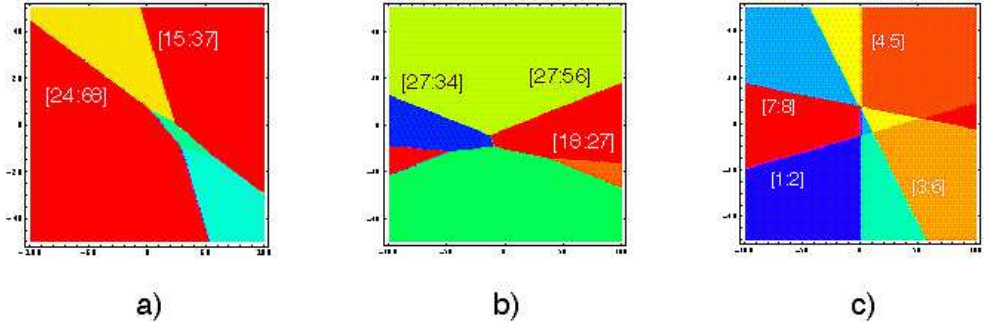}
\caption{Four A-solitons obtained from the $B$-matrix of two D-solitons or three D-solitons.
Two D-solitons in a) is given by the $B$-martrix with ${\bf i}=(1,2,3,6)$ and ${\bf j}=(5,4,7,8)$.
Three D-solitons in b) is given by the $B$-matrix with ${\bf i}=(1,2,3,5)$ and ${\bf j}=(8,7,4,6)$.
Both cases have the same ${\bf p}=(-4,-3,-2,-1,1,3,4,6)$. The four A-soliton solution in c)
is given by the $B$-matrix (\ref{1235B}). Note that the $B$-matrix of (\ref{1236B}) gives the same
solution. }
\label{A8:fig}
\end{figure}

For the same 4-solitons as the previous example, i.e. $[1:2],[3:6],[4:5]$ and $[7:8]$, one can also have the following $B$-matrix
which gives the same $\tau_2$ function,
\begin{equation}\label{1236B}
B=\begin{pmatrix}
0 & 0 & 0 & 1 & 1 & 0 & 0 &0\\
   & 0 & 0 & 1 & 1 & 0 & 0 &0\\
   &   & 0 & 0 & 0 & 0 & -1 & -1\\
   &   &   & 0 & 0 & 0 & 0 & 0\\
  &   &   &   & 0 & 0 & 0 & 0\\
  &   &   &   &   & 0 & 1 & 1\\
  &   &   &   &   &   & 0 & 0\\
  &   &   &   &   &   &  & 0\\
\end{pmatrix}\,.
\end{equation}
This matrix can be constructed from the $B$-matrix for ${\bf i}=(1,2,3,6)$ and ${\bf j}=(5,4,7,8)$,
the case (c) in Proposition \ref{2Dsolitons}, (i.e. this gives two D-solitons with $[1,5:2,4]$ and $[3,7:6,8]$).
In Figure \ref{A8:fig}, we show the example of four A-solitons based on two $B$-matrices:
The case a) corresponds to the $B$-matrix 
for ${\bf i}=(1,2,3,6), {\bf j}=(5,4,7,8)$, and the case b) for ${\bf i}=(1,2,3,5), {\bf j}=(8,7,4,6)$.
We take ${\bf p}=(-4,-3,-2,-1,1,3,4,6)$ for both cases.

\begin{figure}[t]
\includegraphics[width=9cm]{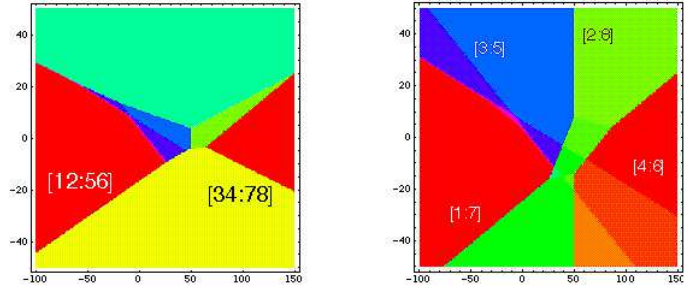}
\caption{The 4-solutions in the right figure is obtained from the $B$-matrices given by (\ref{4BB}).
The left figure shows two D-solitons obtained by the same matrix except $b_{1,7}=0,b_{2,8}=0,b_{3,5}=0$ and $b_{4,6}=0$ (note the extra resonant structure at the intesection region of those D-solitons.}
\label{2-4sol:fig}
\end{figure}

As an example of four A-solitons, we consider the case with $[1:7],[2:8],[3:5]$ and $[4:6]$,
i.e. there are two partial overlaps (two solitons of T-type). 
We first note that this set of 4-solitons can be split into two groups of 2-solitons having no partial overlaps,
$\{[1:7],[2:8]\}$ and $\{[3:5],[4:6]\}$. Since they are both TD-type, we have the TD-type diagram
(up-side-down staircase). This implies we have nonzero entries $b_{1,8},b_{1,2},b_{2,7},b_{7,8}$
from the set $\{[1:7],[2:8]\}$, and $b_{3,4},b_{3,6},b_{4,5},b_{5,6}$ from the second set.
However this $B$-matrix does not give the 4-solitons.
In Figure \ref{2-4sol:fig}, the left figure shows the solution given by this $B$-matrix,
two D-solitons, $[1,2:5,6]$ and $[3,4:7,8]$ with some additional resonances at the
intersection region of those solitons. This can be explained as follows:
This $B$-matrix can be considered as a deformation (bifurcation) from
the $B$-matrix of two D-solitons, $[1,2:5,6]$ and $[3,4:7,8]$, and the $\tau$-function with this $B$-matrix gives resonant
D-solitons, $[1,2:3,4],[1,2:7,8],[3,4:5,6]$ and $[5,6:7,8]$. Then TD-type of 2-solitons
$[1:7],[2:8]$ is obtained from the bifurcation of $[1,2:7,8]$-soliton, which is one of
the resonant solitons. Also note that the exponentials associated to $(1,2,7,8)$ and $(3,4,5,6)$
 become dominant for two D-solitons for either $y\to\infty$ or $y\to-\infty$.
 However those exponentials should not appear for this set of four A-solitons, that is,
 those four A-solitons cannot be obtained from this $B$-matrix.
 This can be fixed by taking following $B$-matrix (i.e. change TD-type to T-type),
 \begin{equation}\label{4BB}
B=\begin{pmatrix}
0 & 1 & 0 & 0 & 0 & 0 & 2 &1\\
   & 0 & 0 & 0 & 0 & 0 & 1 &1\\
   &   & 0 & 1 & 2 & 1 & 0 & 0\\
   &   &   & 0 & 1 & 1 & 0 & 0\\
  &   &   &   & 0 & 1 & 0 & 0\\
  &   &   &   &   & 0 & 0 & 0\\
  &   &   &   &   &   & 0 & 1\\
  &   &   &   &   &   &  & 0\\
  \end{pmatrix}\,.
  \end{equation}
 In Figure \ref{2-4sol:fig}, the right figure shows the 4-solitons generated by this
 $B$-matrix.

We also mention that the $\tau_2$-function with the generic $8\times 8$ $B$-matrix
having all nonzero entries
gives four A-solitons of T-type, that is, they are the same type of the 4-soliton
solution of the KP equation. Just notice that the number of exponential terms
in $\tau_2$ is $\binom{8}{4}=70$, which is the same as the $\tau$-function
for the KP equation (see \cite{biondini:03}). Then the asymptotic analysis
of the $\tau$-functions shows that four A-solitons are given by
$[1:5], [2:6], [3:7]$ and $[4:8]$, which are of T-type.  An example of the generic $B$-matrix
for nonsingular $\tau$-function is given by the skew-symmetric matrix having
1's in all the entries in the upper triangular part.

\section{Multi-soliton solutions}\label{multisolitons}
 For the general case with arbitrary $M=4N$, we consider the $B$-matrix
 with $2N$ numbers of nozero entries $b_{i_k,j_k}$ for $k=1,\ldots,N$
 and $1=i_1<\cdots<i_N,~i_k<j_k$ in the upper triangular part.
 Although there is a scheme to find the $B$-matrix for sign-definite
 $\tau_N$ (i.e. having all the same sign in the pfaffian coefficients),
the generalization of Proposition \ref{2Dsolitons} seems to be difficult.
However, we can show the following Propositions regarding to
 the number of D-solitons for two special cases:
\begin{Proposition}\label{multi-n}
The $\tau_N$-function (\ref{tau}) with $4N\times 4N$ $B$-matrix given by $J_0$ of (\ref{J0})
can have only $N$ D-solitons in the generic situation.
\end{Proposition}
\begin{Proof}
Apply Lemma \ref{cij}, and find the dominant exponentials by following
the argument of ``level of intersection'' introduced in \cite{biondini:03} (the main idea here is to find the dominant exponents from the graphs of $u_{i,j}(c)$).
\end{Proof}

\begin{figure}[t]
\includegraphics[width=12cm]{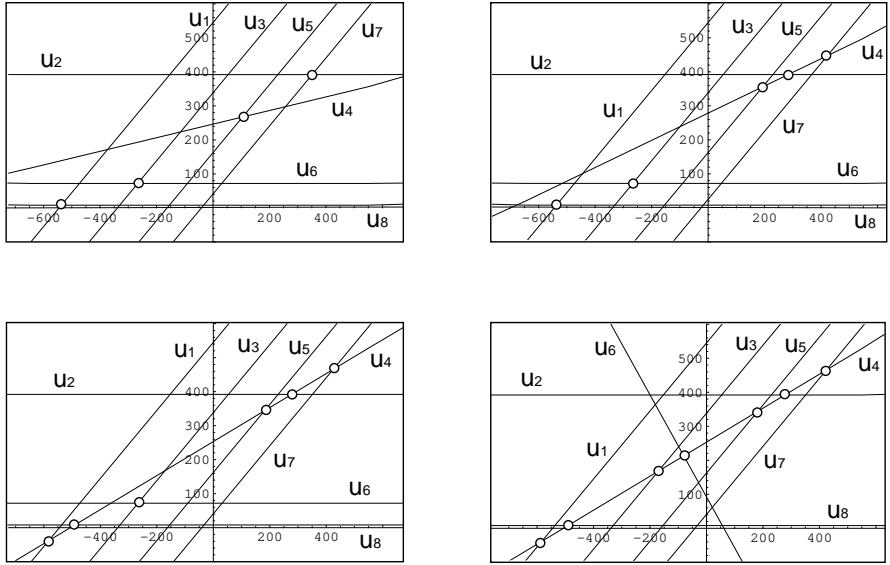}
\caption{The graphs of $u_k(c)=(p_{k}+p_{17-k})c+p_k^2+p_{17-k}^2$ for $k=1,\ldots,16$.
Each open circle of the intersections of two functions $u_k$ corresponds to a D-soliton
given by $\tau_4$. For eaxmple,
  the point for $u_1=u_8$ corresponds to the D-soliton of $[1,16:8,9]$.
  The number of open circles depends on the choice of the parameter ${\bf p}=(p_1,\ldots,p_{16})$.
  Notice that only $u_4$ and later $u_6$ are changed to increase the number of solitons 
  (see the text for the explicit values of ${\bf p}$). }
\label{M16:fig}
\end{figure}

The other case corresponds to the $B$-matrix, $B=J_\pi$, where 
$J_\pi$ is defined by $J_\pi:=\pi J_0\pi^T$ with $\pi\in S_{4N}$, 
the permutation given by $\pi=[2:4N]\cdot[4:4N-2]\cdots[2N:2N+2]$, i.e.
\begin{equation}\label{Jw}
J_w=\begin{pmatrix}
0 & 0 & \cdots & 0         & 0         &\cdots  & 0           &1\\
   & 0 & \cdots & 0         & 0          & \cdots & 1          &0\\
   &   & \ddots & \vdots & \vdots & \vdots & \vdots & \vdots \\
   &   &             & 0           & 1         & \cdots  &0         & 0\\
  &   &              &             & 0           & \cdots          & 0          & 0\\
  &   &               &             &           & \ddots & \vdots  &\vdots\\
  &   &               &            &            &             & 0           & 0\\
  &   &               &            &            &             &               & 0\\
  \end{pmatrix}
 \end{equation}
 Then we have:
\begin{Proposition}\label{multi-a}
The number of D-solitons generated by 
the $\tau_N$-function (\ref{tau}) with $4N\times 4N$ $B$-matrix given by $J_w$ of (\ref{Jw})
can be any number from $N$ to $2N-1$, depending on the choice of the parameter ${\bf p}$.
\end{Proposition}
Here instead of giving a precise proof, we demonstrate 
a case with $M=16~(N=4)$: In Figure \ref{M16:fig}, we show the graphs of
$u_k(c)=(p_k+p_{17-k})c+p_k^2+p_{17-k}^2$ for $k=1,\ldots,8$.
With a proper choice of the parameter ${\bf p}=(p_1,\ldots,p_{16})$,
one can find four different cases of intersection patterns of those lines.
Then D-solitons can be found at the intersection points marked by open circles in Figure \ref{M16:fig}, that is, the level of intersection of those points should be four (i.e. there are four
lines above those points). For example, D-soliton $[4,13:5,12]$ is obtained by the
intersection of $u_4$ and $u_5$, and two dominant exponents for this soliton are
$\hat E_1\hat E_2\hat E_3\hat E_4$ and $\hat E_1\hat E_2\hat E_3\hat E_5$ which are
obtained from four lines above the intersection point $u_4=u_5$ (this is the definition of
the level of intersection; see \cite{biondini:03} for more details).
We have the following four cases:
\begin{itemize}
\item[(a)] Four D-solitons (the top left figure) of $[1,16:8,9], [2,15:7,10], [3,14:6,11]$ and $[4,13:5,12]$
with the parameter,
\[{\bf p}=(-16,-14,-12.5,-11,-8.5,-6,-4,-2,2,5,6,9.5,11.2,13.5,14,17)\,.\]
\item[(b)] Five D-solitons (the top right figure) of $[1,16:8,9], [2,15:4,13], [3,14:6,11], [4,13:5,12]$ and $[4,13:7,10]$ with
\[ {\bf p}=(-16,-14,-12.5,-11.6,-8.5,-6,-3,-2,2,4,6,9.5,12,13.5,14,17)\]
Note here that three D-solitons have the common index $(4,13)$. The $[2,15:7,10]$-soliton
in the case (a) causes the resonant interaction with two D-solitons $[2,15:4,13]$ and
$[4,13:7,10]$.
\item[(c)] Six D-solitons (the bottom left figure) of
$[1,16:4:13], [2,15:4,13], [3,14:6,11], [4,13:5,12], [4,13:7,10]$ and $[4,13:8,9]$ with
\[ {\bf p}=(-16,-14,-12.5,-11,-8.5,-6,-4,-2,2,5,6,9.5,11.5,13.5,14,17)\,.\]
\item[(d)] Seven D-solitons (the bottom right figure) of
$[k,17-k:4,13]$ for all $k=1,\ldots,8$, except $k=4$, and
\[ {\bf p}=(-16,-14,-12.5,-11,-8.5,-7.5,-4,-2,2,5,6,9.5,11.5,13.5,14,17)\,.\]
\end{itemize}
One should note that those ${\bf p}$ values do {\it not} give generic cases, since some of the lines $u_k(c)$
are parallel as shown in Figure \ref{M16:fig}. However, one can make those to be generic by changing slightly
the parameters $p_i$'s without breaking the intersection patterns. (We show those non-generic parameters, since those provide a clear evidence of the existence of those solitons.)

We also remark that all the $N$-soliton solutions obtaind by $\tau_N$ with the $B$-matrix 
given in Proposition \ref{multi-n} are fully resonant cases
like the T-type for the KP equation, that is, the $\tau_N$ contains $\binom{2N}{N}$ exponential terms
having the form $\prod_{k=1}^N{\hat E}_{k}=\prod_{k=1}^NE_{i_k,j_k}$ (see \cite{biondini:03}).

Finally, we note that in the case of generic $B$-matrix having all nonzero entries, the $\tau_N$ function has the maximal number 
of exponential terms $\binom{4N}{2N}$ which has the same structure of the $\tau$-function
for fully resonant $2N$-soliton solutions (i.e. T-type) of the KP equation. Those $2N$ solitons
are all A-type, and the interaction patterns are topologically the same for both KP and DKP equations.
In those cases, the functions $v^{\pm}$ appear only locally at the interaction regions among $2N$-solitons.

\bibliographystyle{amsalpha}

\end{document}